\documentclass[letterpaper]{article}
\usepackage{aaai25}
\usepackage{times}
\usepackage{helvet}
\usepackage{courier}
\usepackage[hyphens]{url}
\usepackage{graphicx}
\usepackage{natbib}
\usepackage{caption}
\usepackage{algorithm}
\usepackage{algorithmic}
\usepackage{booktabs}
\usepackage{array}
\usepackage{ragged2e}
\usepackage{placeins}
\usepackage{float}
\newcolumntype{P}[1]{>{\RaggedRight\arraybackslash}p{#1}}
\usepackage{newfloat}
\usepackage{listings}
\usepackage{subcaption}

\usepackage{enumitem}
\usepackage{tcolorbox}
\tcbuselibrary{breakable}
\usepackage{xcolor}
\usepackage{tabularx}
\usepackage{multirow}
\usepackage[titletoc,title]{appendix}

\DeclareCaptionStyle{ruled}{labelfont=normalfont,labelsep=colon,strut=off}
\floatstyle{ruled}
\newfloat{listing}{tb}{lst}{}
\floatname{listing}{Listing}
\newcommand{\answerYes}[1]{\textbf{Yes.} #1}

\newcommand{\answerNA}[1]{\textbf{N/A.} #1}

\title{Grievance Politics vs. Policy Debates: A Cross-Platform Analysis of Conservative Discourse on Truth Social and Reddit}

\author{
    Yining Wang\textsuperscript{\rm 1,2},
    Alhasan Abdellatif\textsuperscript{\rm 3},
    Artemis Deligianni\textsuperscript{\rm 4},
    Hannah Hok\textsuperscript{\rm 5,6},
    Yusuf M\"ucahit \c{C}etinkaya\textsuperscript{\rm 4},
    Tu\u{g}rulcan Elmas\textsuperscript{\rm 4}
}
\affiliations{
    \textsuperscript{\rm 1}GESIS -- Leibniz Institute for the Social Sciences\\
    \textsuperscript{\rm 2}RWTH Aachen University\\
    \textsuperscript{\rm 3}Heriot-Watt University\\
    \textsuperscript{\rm 4}The University of Edinburgh\\
    \textsuperscript{\rm 5}Massachusetts Institute of Technology\\
    \textsuperscript{\rm 6}Harvard University\\
    yining.wang@gesis.org, alhasanabdellatif@gmail.com, A.Deligianni@sms.ed.ac.uk, hok1@mit.edu,  v1ymucah@ed.ac.uk, tugrulcan@ed.ac.uk
}

\usepackage{bibentry}

\begin{document}

\maketitle

\begin{abstract}

We present the first large-scale comparative analysis of Truth Social and the most popular conservative Reddit communities, r/Conservative, r/conservatives, and r/Republican. Using topic modeling with FASTopic and LLM-assisted refinement, we analyze topic prevalence, toxicity, and temporal dynamics across these communities during the first eight months of Truth Social. We find clear contrasts: Truth Social centers on grievance and narrative-driven content, while Reddit focuses more on policy debates. Toxicity is higher on Reddit and peaks in cultural and leader-focused topics. Despite similar event-driven participation shocks across platforms, Truth Social shows higher baseline proportions of users engaging with political topics. Our findings contribute to understanding how alternative right-leaning platforms reshape online discourse.

\end{abstract}

\section{Introduction}
The social media ecosystem is increasingly fractured by debates over content moderation, algorithmic bias, and political fairness on major platforms. In the U.S., conservative users and elites have frequently accused major platforms of suppressing right-leaning viewpoints~\cite{gillespie2018custodians}, a narrative amplified by major deplatforming events such as the widespread removal of QAnon-related content~\cite{monti2023online}. This led to the emergence of alternative social media platforms such as Parler, Gab, and Truth Social that are designed to target right-wing audiences under the banner of ``free speech.'' These platforms are often criticized for fostering echo chambers, fringe ideologies, and the spread of conspiratorial narratives~\cite{kristensen2025platform}.
The potential for such online dynamics to translate into harmful real-world consequences makes the study of these spaces a pressing issue for understanding contemporary political communication. In this study, we focus on Truth Social, one of the most prominent alternative right-leaning social media platforms, founded in October 2021 by Donald Trump. Its close association with Trump distinguishes it from other alternative platforms and motivates a comparison with more established conservative forums, including those on Reddit.

From a political communication perspective, online platforms can serve distinct communicative roles within an ideological ecosystem. Prior work distinguishes between grievance-oriented political communication, characterized by affective appeals, leader-centered narratives, and institutional distrust, often associated with populist and conspiratorial styles of politics \cite{uscinski2019conspiracy, moffitt2020populism, armaly2025disparate}, and policy-oriented discourse, which emphasizes issue-specific debate, ideological boundary-drawing, and argumentative engagement \cite{habermas1991structural, Grimmer_Stewart_2013}.

Building on theories of platform affordances and political communication, we conceptualize Truth Social and conservative Reddit communities as structurally distinct environments that may privilege different communicative forms. Platforms designed around elite amplification may facilitate grievance-based and personalized narratives, while more structured discussion spaces with active moderation may channel political conflict into policy debates and ideological contestation. To provide an empirical baseline for evaluating these differences, we compare Truth Social with the three biggest conservative subreddits: \textit{r/Republican}, \textit{r/Conservative}, and \textit{r/conservatives}. Our research questions are:

\begin{enumerate}[label=\textbf{{RQ{\theenumi}}},leftmargin=26pt]
\item Which political topics dominate conservative discourse on Truth Social and Reddit, and what do their differences reveal about each platform’s role in the conservative information ecosystem?

\item How do the level and form of toxic speech vary across platforms and topics?

\end{enumerate}

Our main contribution is the first large-scale analysis of content on Truth Social. In doing so, we also present the first comparative study of Truth Social and conservative Reddit communities, characterizing their linguistic and ideological landscapes in a unified framework. Methodologically, we experiment and show that FASTopic outperforms popular topic modeling techniques LDA and BERTopic for this task.

By analyzing topics, we find that Truth Social disproportionately amplifies and sustains a persistent focus on grievance-oriented and leader-centered narratives. Meanwhile, Reddit communities emphasize policy debates and cultural controversies. Toxicity is concentrated in cultural and leader-related topics across platforms, with Reddit exhibiting higher baseline levels. Our findings contribute to research on alternative platforms, political polarization, and online grievance politics by demonstrating that alternative right-leaning platforms do not merely amplify toxicity, but reorganize political discourse around different modes of political communication.

\section{Related Work}

This study builds on research on platform moderation and deplatforming, user migration to alternative platforms, echo chambers, and computational analyses of political discourse.

\noindent\textbf{Platform Moderation and Deplatforming:} Content moderation has increasingly been conceptualized as a form of platform governance rather than a purely technical intervention \cite{gillespie2018custodians, vega2021online}. Decisions about what content is allowed, removed, or de-prioritized shape participation, visibility, and power relations within online spaces, making moderation a central political and normative issue. Recent systematization efforts highlight the complexity of this governance landscape, documenting how moderation practices vary across platforms in guidelines, enforcement mechanisms, and outcomes \cite{singhal2023sok}. Across political contexts, moderation practices have been criticized as illegitimate or biased, with users from diverse ideological positions reporting perceptions of censorship and unfair treatment~\cite{haimson2021disproportionate,ccetinkaya2025state}.

High-profile de-platforming events have intensified these debates. The banning of Donald Trump from Twitter and Facebook following the January 6, 2021 Capitol riot \cite{alizadeh2022content} exemplifies how moderation actions can trigger political backlash and reconfigure online ecosystems. In response, a growing number of alternative or ``alt-tech'' platforms have emerged, explicitly branding themselves as pro--free speech alternatives to mainstream social media \cite{failla2024m, kolluri2025quantifying}. Although these platforms present themselves as neutral spaces for unrestricted expression, prior research shows that reduced moderation is associated with greater ideological insulation, conspiratorial discourse, and higher toxicity \cite{kumarswamy2025causal}. Furthermore, some research has suggested that de-platforming rarely eliminates ideological communities, which often reconstitute on alternative platforms while preserving shared narratives instead. Truth Social’s direct association with Donald Trump highlights how leader-centered platforms may function differently from decentralized conservative spaces \cite{hughes2025echo, bidewell2026gendered}.

\noindent\textbf{User Migration and Platform Reconfiguration:} Prior work shows that users migrate across platforms in response to moderation actions, policy changes, and shifting platform norms. Drawing on push--pull theories, studies identify perceived censorship and governance dissatisfaction as push factors, and ideological alignment and permissive moderation as key pull factors \cite{fiesler2020moving}.

Empirical studies show that moderation interventions trigger measurable user migration, with many displaced users relocating to ideologically adjacent spaces on the same or alternative platforms \cite{horta2021platform}. Conspiracy-oriented communities exhibit even greater resilience, as deplatformed QAnon groups often migrated wholesale to fringe platforms such as Voat, where they reassembled social networks and sustained high activity \cite{monti2023online}.

Large-scale de-platforming following January 2021 highlights cross-platform migration dynamics, with toxicity declining on mainstream platforms but increasing on alternative platforms as displaced users regrouped \cite{buntain2023cross}. These patterns suggest that de-platforming reduces visibility in mainstream spaces while concentrating ideological and toxic content elsewhere.

More recent studies highlight migration driven by platform instability or ownership changes rather than explicit bans. User movements from Twitter to platforms such as Mastodon, BlueSky, and Threads illustrate how governance uncertainty and perceived norm shifts can catalyze migration even without direct moderation actions \cite{failla2024m}.

\noindent\textbf{Echo Chambers and Conspiratorial Discourse:} The dynamics of platform moderation and user migration intersect closely with echo chambers and polarization. literature demonstrates that ideologically homogeneous environments reinforce prior beliefs, reduce exposure to cross-cutting perspectives, and increase affective and ideological polarization \cite{bright2016explaining, cinelli2021echo}. Systematic reviews further identify partisan sorting and ideological extremism as key drivers of insular online discourse \cite{hartmann2025systematic, ccetinkaya2025narra}. Empirical studies show that right-leaning echo chambers tend to exhibit higher cohesion and isolation than left-leaning ones \cite{ccetinkaya2025cross}, particularly during periods of political crisis such as the COVID-19 pandemic \cite{jiang2021social}. These findings suggest that migration to alternative platforms with permissive moderation may intensify ideological homogeneity and grievance-oriented discourse. Research on conspiratorial thinking further contextualizes these patterns. Individuals at ideological extremes are more prone to conspiracy beliefs than moderates \cite{van2015voters, imhoff2022conspiracy}, and in the U.S. context, conservatives are more likely to endorse narratives that delegitimize institutions, media, or political opponents \cite{enders2023republicans}. Such beliefs are closely linked to populism, distrust of elites, and polarized worldviews \cite{uscinski2022psychological}, making alternative platforms fertile ground for grievance-centered and conspiratorial narratives.

\noindent\textbf{Computational Approaches on Political Discourse:} Toxicity and incivility are widely used indicators of discursive quality in online political environments. Prior work consistently shows that hostile language clusters around identity-charged and highly partisan topics \cite{jhaver2018online, pavlopoulos-etal-2020-toxicity}, and that moderation intensity and platform design strongly shape levels of aggression, with weaker enforcement associated with higher toxicity \cite{10.1145/3359265}. Studies of user migration further suggest that de-platforming may reduce harmful content on mainstream platforms while increasing toxicity on alternative platforms as displaced users regroup in ideologically homogeneous spaces \cite{horta2021platform, buntain2023cross}.
Our study builds on computational approaches for large-scale analysis of political discourse, particularly topic modeling for uncovering thematic structure \cite{Grimmer_Stewart_2013}. Recent LLM-assisted methods improve interpretability and reduce manual labeling, enabling more reliable cross-platform comparisons~\cite{yang2025llm}.

\section{Data \& Methodology}
We outline our data sources and preprocessing pipeline, followed by computational methods used to analyze political discourse across platforms.

\noindent\textbf{Data Collection:} Our study uses data from Truth Social and Reddit. For Truth Social, we employ two datasets with distinct collection methodologies to ensure robustness. The first is a public dataset by Gerard et al.~\citeyear{gerard2023truth} which we name \emph{Truth Social Core}. It spans from Truth Social's launch on February 21, 2022 to October 15, 2022 and consists of 454,458 users and 823,927 posts authored by 17,245 users. The posts include original posts ("Truths") (305,356), replies (279,755), and reposts ("ReTruths") (238,841). Gerard et al. employ snowball sampling to collect this data. They first collect posts of followers of \textbf{\textit{@realDonaldTrump}} and expand through breadth-first crawling that navigates these users' followers. This collection method introduces a bias toward \textbf{\textit{@realDonaldTrump}} and politically engaged users within his orbit during collection.

To mitigate this bias, we construct a second Truth Social dataset named \emph{Truth Social Extra} using a different collection strategy. We employ the public dataset by Shah et al.~\citeyear{shah2024unfiltered} which is collected by searching for posts containing keywords related to the 2024 U.S. election, rather than follower based crawling.

From the 50,459 users in this dataset, we identify users whose posted prior to October 2022 (matching our study period) and whose was not already collected by Gerard et al~\cite{gerard2023truth}. We randomly sample 2524 of these unique users and collect their posts over the same time frame as the \emph{Truth Social Core} using TruthBrush. This dataset comprises 186,195 posts (``Truths"), including 109,845 reposts (``ReTruths”). Crucially, \emph{Truth Social Extra} represents a distinct user base that did not follow Donald Trump or his extended network in October 2022, suggesting these users may exhibit behavioral patterns distinct from those in the core Trump orbit.

To capture conservative discourse on Reddit, we download Reddit data of 2022 provided by Project Arctic Shift~\cite{arcticshift2022}. We identify the largest conservative subreddits in this dataset that maintain their size to date: r/Conservative (1.3M members), r/Republican (216K), and r/conservatives (114K). We collect posts and comments from these three subreddits over the same period as Truth Social data. We exclude smaller subreddits with narrower focus such as r/trump (70K members) and r/Libertarian (56k members) to ensure broader representation. While this limits the sample to political subreddits, it provides a focused proxy for right-wing discourse comparison.

\noindent\textbf{Preprocessing:} To ensure data quality and comparability across platforms, we apply consistent preprocessing steps. Our goal is to isolate posts with substantive textual content. We remove posts that contain only URLs, emojis, or user mentions, as they provide little meaningful information. We also eliminate duplicate entries to avoid redundancy and filter out posts with invalid timestamps, those with missing values or falling outside the study timeframe (February 21 - October 15, 2022), to maintain chronological consistency.

Finally, to keep the analysis linguistically consistent, we applied language detection to the full dataset using the fastText language identification model. The results show that 99.0\% of the posts are in English. This is expected, as both Truth Social and the selected Reddit communities mainly serve U.S.-based, English-speaking users. We keep all posts in the analysis because the small number of non-English posts does not meaningfully affect the results.

Following preprocessing, we remove around 10\% of the posts from r/conservatives, 25\% from r/Republican, 30\% from r/Conservative, 43\% from \emph{Truth Social Core} and 34\% from \emph{Truth Social Extra}. The higher removal rates on certain platforms reflect differences in posting behavior; Truth Social, in particular, contain a larger proportion of URL-only shares and reposts without added commentary. Notably, filtering invalid timestamps, posts with empty timestamps, reduced the number of users from 17,245 to 2,695 in \emph{Truth Social Core}, and from 3065 to 1000.

\noindent\textbf{Dataset Summary:} Table \ref{tab:post_summary} summarizes dataset statistics following preprocessing. Notably, \textit{Truth Social Core} is comparable in total volume to \textit{r/Conservative} (476k versus 499k posts), yet this content is sourced from a significantly smaller user base (2,695 versus 65,917 users). This disparity results in a much higher average engagement rate on the newer platform: Truth Social users in our datasets average over 120 posts per user, whereas users average fewer than 8 in conservative Reddit communities.

\begin{figure}[ht]
    \centering
    \includegraphics[width=\columnwidth]{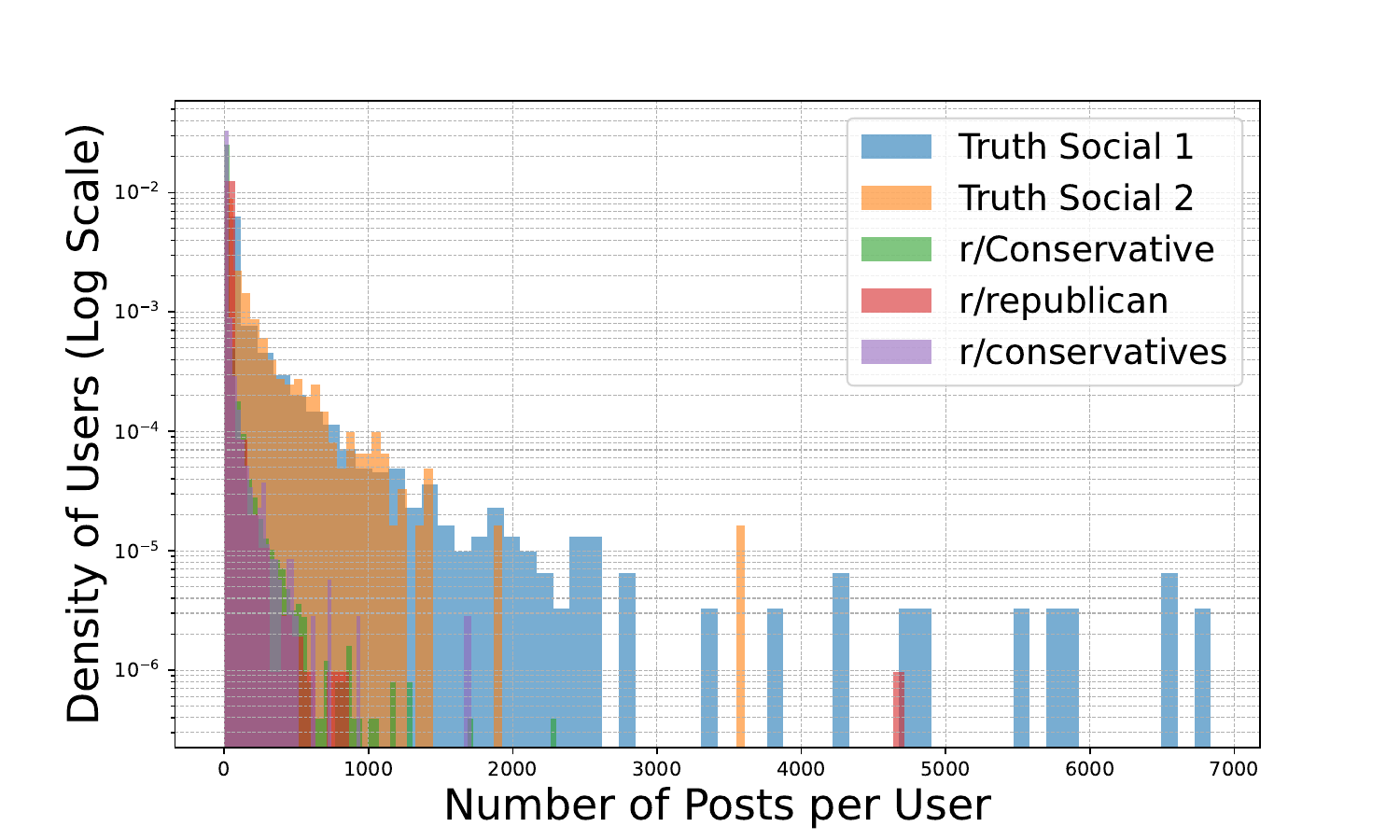}
      \caption{Distribution of posts per user across Truth Social and conservative Reddit communities (logarithmic scale).}
        \label{fig:post_distributions}
\end{figure}

Figure \ref{fig:post_distributions} shows the histogram of users by post count. The distribution is highly skewed for all datasets, a pattern typical of online social media engagement: most users contribute only a small number of posts, while a small minority account for a disproportionately large share of content. Truth Social displays a notably heavier tail with a much larger proportion of users contributing at least 1000 posts, unlike Reddit where very few users exceed this threshold.

These differences may be partially attributed to our sampling methodology, as we strictly collect Reddit activity within specific subreddits (potentially missing users' posts elsewhere), whereas Truth Social data reflects broader user timelines. However, they may also indicate distinct platform cultures or design features that encourage hyper-active posting behavior, a factor requiring further investigation.

\begin{table}[ht]
\centering
\footnotesize
\begin{tabular}{lrrrr}
Platform & \multicolumn{1}{c}{Posts} & \multicolumn{1}{c}{Users} & \multicolumn{1}{c}{Posts/User} \\
\midrule
Truth Social Core   & 476,119 & 2,695  & 176.67 \\
Truth Social Extra   & 123,818  & 1,000  & 123.82 \\
r/Conservative  & 498,720 & 65,917 & 7.57   \\
r/Republican    & 90,303  & 13,228 & 6.83   \\
r/conservatives & 97,701  & 12,232 & 7.99   \\
\end{tabular}
\caption{Datasets statistics upon preprocessing.}
\label{tab:post_summary}
\end{table}

Although this study focuses on the eight-month window within Truth Social's launch in February 2022, our analysis period captures a platform that had already reached operational maturity. Figure~\ref{fig:daily_counts} (Appendix A) shows that the daily activity stabilizes after April 2022 and Figure~\ref{fig:daily_counts_cum} (Appendix A) shows the majority of users in both Truth Social datasets had posted at least once by May 2022. By June 2022, the user bases of both Truth Social Core and Extra reach relatively stable plateau levels. This suggests that our analysis captures the platform during a steady-state period rather than merely its chaotic launch phase, lending confidence that the behavioral patterns we observe reflect established community norms rather than early-adoption dynamics.

\subsection{Data Labeling}
We use a multi-stage labeling process with established language models to identify political content and toxicity.

\noindent\textbf{Political Content Filtering:} To focus on political discourse and reduce noise from irrelevant posts, we first filter out non-political content using a zero-shot classification model, BART-Large-MNLI \cite{lewis2019bart,yin2019benchmarking}, to classify posts as political or non-political. The model, fine-tuned on the Multi-Genre Natural Language Inference dataset, can classify text into predefined categories without task-specific training.

To evaluate the classification and select a confidence threshold, we create a test set by manually annotating a random sample of 400 posts. Two annotators independently annotate the posts and reported substantial agreement (Cohen's $\kappa$ = 0.65)~\cite{cohen1960coefficient}. As Table \ref{tab:threshold_evaluation} shows, the model with a threshold of 0.7 achieves the highest accuracy, 74.8\%, and near-optimal F1 of 0.784. We choose this threshold as it offers better precision than the 0.6 threshold.

The share of political content is 52\% on \emph{Truth Social Core}, 65\% on \emph{Truth Social Extra}, 61\% on r/Republican, 58\% on r/Conservative, and 60\% on r/conservatives. We only include these posts in the analysis.

\begin{table}[h!]
\centering
\resizebox{0.8\linewidth}{!}{
\begin{tabular}{ccccc}
\textbf{Threshold} & \textbf{Accuracy} & \textbf{Recall} & \textbf{Precision} & \textbf{F1} \\
\midrule
0.1 & 0.580 & 0.983 & 0.582 & 0.731 \\
0.2 & 0.612 & 0.970 & 0.603 & 0.744 \\
0.3 & 0.628 & 0.948 & 0.616 & 0.747 \\
0.4 & 0.682 & 0.927 & 0.662 & 0.772 \\
0.5 & 0.702 & 0.884 & 0.690 & 0.775 \\
0.6 & 0.738 & 0.828 & 0.747 & 0.785 \\
\textbf{0.7} & \textbf{0.748} & \textbf{0.789} & \textbf{0.779} & \textbf{0.784} \\
0.8 & 0.728 & 0.668 & 0.829 & 0.740 \\
0.9 & 0.682 & 0.534 & 0.867 & 0.661 \\
\end{tabular}
}
\caption{Political content classification results}
\label{tab:threshold_evaluation}
\end{table}

\noindent\textbf{Toxicity:} We measure toxicity in the discourse using the Perspective API~\cite{jigsaw2025perspective}. The API employs machine learning models trained on millions of human-annotated comments to identify potentially harmful content, with scores ranging from 0 (least toxic) to 1 (most toxic). For this study, we select five dimensions that capture distinct aspects of toxicity: \texttt{TOXICITY} (overall rudeness), \texttt{INSULT} (personal attacks), \texttt{THREAT} (violent language), \texttt{SEVERE TOXICITY} (extremely harsh content), and \texttt{IDENTITY ATTACK} (negative generalizations about identity groups).

We acknowledge that the Perspective API has known limitations, including potential biases related to dialect and identity terms~\cite{sap2019risk, hutchinson2020social}. Despite these limitations, we choose Perspective API for three reasons. First, it is a widely used and well-documented tool for toxicity detection, which facilitates comparison with prior studies~\cite{hickey2025x}. Second, it provides consistent, multi-dimensional toxicity scores at scale, making it suitable for large cross-platform analyses. Third, our analysis focuses on \textit{relative differences} between platforms and topics rather than absolute toxicity levels. Since the same classifier is applied uniformly across all datasets, any systematic bias is shared across platforms, preserving the validity of comparative patterns. This approach aligns with prior work that uses Perspective API for relative toxicity analysis across online communities~\cite{rajadesingan2020quick} rather than absolute toxicity levels. To further mitigate potential bias, we interpret results at the aggregate level and avoid drawing conclusions about individuals.

Since scores of 0.5 and under indicate a text is not toxic~\cite{kumarswamy2025causal}, we conduct a robustness check using a thresholding approach, where values $\leq 0.5$ were set to zero. This transformation did not affect the relative differences between sources or alter our substantive conclusions. Consequently, we report results using the original continuous scores.
To assess if there are meaningful differences in the amount of toxic posts across platforms we randomly sample 4000 observations from each and we fit a Bayesian Logistic regression with user random effects to account for multiple observations per user.

    \label{fig:toxicity_radar}

\subsection{Topic Modeling}
We aim to analyze large-scale conservative discourse across platforms, where posts span different political issues. Given the scale and heterogeneity of data, topic modeling is well suited for this task. It uncovers recurring themes and organizes discussions into interpretable clusters, allowing us to capture broad patterns rather than isolated examples. To ensure reliable results, we systematically test different models and numbers of topics to choose optimal settings.

We randomly sample 50,000 posts from each community, 250,000 posts in total, to facilitate comparison across platforms and reduce computation overload in topic modeling.

We evaluate three topic modeling approaches: LDA, BERTopic, and FASTopic. Implementation details and hyperparameter settings are provided in Appendix B ~\ref{app:topic_modeling}.

We initially experiment with 5 different topic numbers (\textit{K}). For each topic number, we repeat the experiment five times with different random seeds and report mean scores with standard deviations. We evaluate models using three metrics that capture different aspects of topic quality:

\noindent\textbf{Topic Coherence (CV):} measures how interpretable topics are by quantifying how often their top words co-occur in the reference corpus~\cite{roder2015exploring}. We use the $C_V$ score from the Palmetto toolkit, which combines a sliding window, normalized pointwise mutual information (NPMI), and cosine similarity. For each model, we compute average $C_V$ across topics using the top-10 words. Higher scores indicate more semantically consistent topics.

\noindent\textbf{Topic Diversity (TD):} measures distinctiveness by calculating the proportion of unique words among top-$n$ topic words~\cite{dieng2020topic}. Following prior work, we set $n=10$. A score of 1.0 means no repetition across topics, while lower values indicate redundancy. TD complements coherence by ensuring topics are non-overlapping.

\noindent\textbf{Topic Quality (TQ):} combines the two metrics as $TQ = CV \times TD$~\cite{dieng2020topic}. This captures the trade-off between coherence (within-topic quality) and diversity (across-topic coverage). Higher TQ values indicate topics that are both interpretable and distinct.

Table~\ref{tab:model_comparison} reports results across different numbers of topics ($K$). LDA shows moderate coherence but suffers from low diversity, often recycling the same keywords across topics. BERTopic achieves high diversity but the lowest coherence scores, as its clusters frequently include incoherent or weakly related words. FASTopic consistently achieves best performance across all three metrics, combining highest coherence with strong diversity, resulting in highest topic quality across all values of $K$. Thus, we select FASTopic as the modeling framework for subsequent analysis.

\begin{table}[t]
\centering
\scriptsize
\renewcommand{\arraystretch}{0.85}
\setlength{\tabcolsep}{4pt}
\begin{tabular}{lcccc}
\textbf{K} & \textbf{Model} & \textbf{TC} & \textbf{TD} & \textbf{TQ} \\
\midrule
\multirow{3}{*}{10}
 & LDA      & 0.477\textsubscript{\,\textit{±0.009}} & 0.700\textsubscript{\,\textit{±0.018}} & 0.334\textsubscript{\,\textit{±0.010}} \\
 & BERTopic & 0.372\textsubscript{\,\textit{±0.007}} & 0.805\textsubscript{\,\textit{±0.034}} & 0.299\textsubscript{\,\textit{±0.015}} \\
 & FASTopic & \textbf{0.502}\textsubscript{\,\textit{±0.015}} & \textbf{0.880}\textsubscript{\,\textit{±0.022}} & \textbf{0.441}\textsubscript{\,\textit{±0.018}} \\
\midrule
\multirow{3}{*}{20}
 & LDA      & 0.463\textsubscript{\,\textit{±0.008}} & 0.665\textsubscript{\,\textit{±0.016}} & 0.308\textsubscript{\,\textit{±0.009}} \\
 & BERTopic & 0.380\textsubscript{\,\textit{±0.012}} & 0.845\textsubscript{\,\textit{±0.015}} & 0.321\textsubscript{\,\textit{±0.014}} \\
 & FASTopic & \textbf{0.495}\textsubscript{\,\textit{±0.012}} & \textbf{0.910}\textsubscript{\,\textit{±0.019}} & \textbf{0.451}\textsubscript{\,\textit{±0.016}} \\
\midrule
\multirow{3}{*}{30}
 & LDA      & 0.457\textsubscript{\,\textit{±0.010}} & 0.607\textsubscript{\,\textit{±0.019}} & 0.277\textsubscript{\,\textit{±0.011}} \\
 & BERTopic & 0.382\textsubscript{\,\textit{±0.008}} & 0.862\textsubscript{\,\textit{±0.006}} & 0.329\textsubscript{\,\textit{±0.007}} \\
 & FASTopic & \textbf{0.485}\textsubscript{\,\textit{±0.014}} & \textbf{0.890}\textsubscript{\,\textit{±0.025}} & \textbf{0.432}\textsubscript{\,\textit{±0.020}} \\
\midrule
\multirow{3}{*}{40}
 & LDA      & 0.463\textsubscript{\,\textit{±0.011}} & 0.548\textsubscript{\,\textit{±0.017}} & 0.254\textsubscript{\,\textit{±0.008}} \\
 & BERTopic & 0.384\textsubscript{\,\textit{±0.009}} & 0.865\textsubscript{\,\textit{±0.006}} & 0.332\textsubscript{\,\textit{±0.009}} \\
 & FASTopic & \textbf{0.479}\textsubscript{\,\textit{±0.013}} & \textbf{0.923}\textsubscript{\,\textit{±0.021}} & \textbf{0.442}\textsubscript{\,\textit{±0.017}} \\
\midrule
\multirow{3}{*}{50}
 & LDA      & 0.453\textsubscript{\,\textit{±0.012}} & 0.504\textsubscript{\,\textit{±0.020}} & 0.228\textsubscript{\,\textit{±0.010}} \\
 & BERTopic & 0.383\textsubscript{\,\textit{±0.006}} & 0.868\textsubscript{\,\textit{±0.015}} & 0.332\textsubscript{\,\textit{±0.010}} \\
 & FASTopic & \textbf{0.457}\textsubscript{\,\textit{±0.016}} & \textbf{0.870}\textsubscript{\,\textit{±0.018}} & \textbf{0.397}\textsubscript{\,\textit{±0.019}} \\
\end{tabular}
\caption{Topic model evaluation across topic counts ($K$).}
\label{tab:model_comparison}
\end{table}

\noindent\textbf{Choosing the Number of Topics ($K$):} To determine a suitable number of topics, we train FASTopic with $K$ values ranging from 5 to 100 in increments of 5.

We evaluate models using two complementary metrics: topic quality (TQ), which reflects overall semantic quality of topics, and reconstruction loss, which measures how well the model reproduces original documents from learned topic representations. Lower reconstruction loss means document–topic and topic–word distributions capture data structure more faithfully, while higher loss suggests topics fail to explain the corpus effectively.

\begin{figure}[ht!]
    \centering
    \includegraphics[width=1\columnwidth]{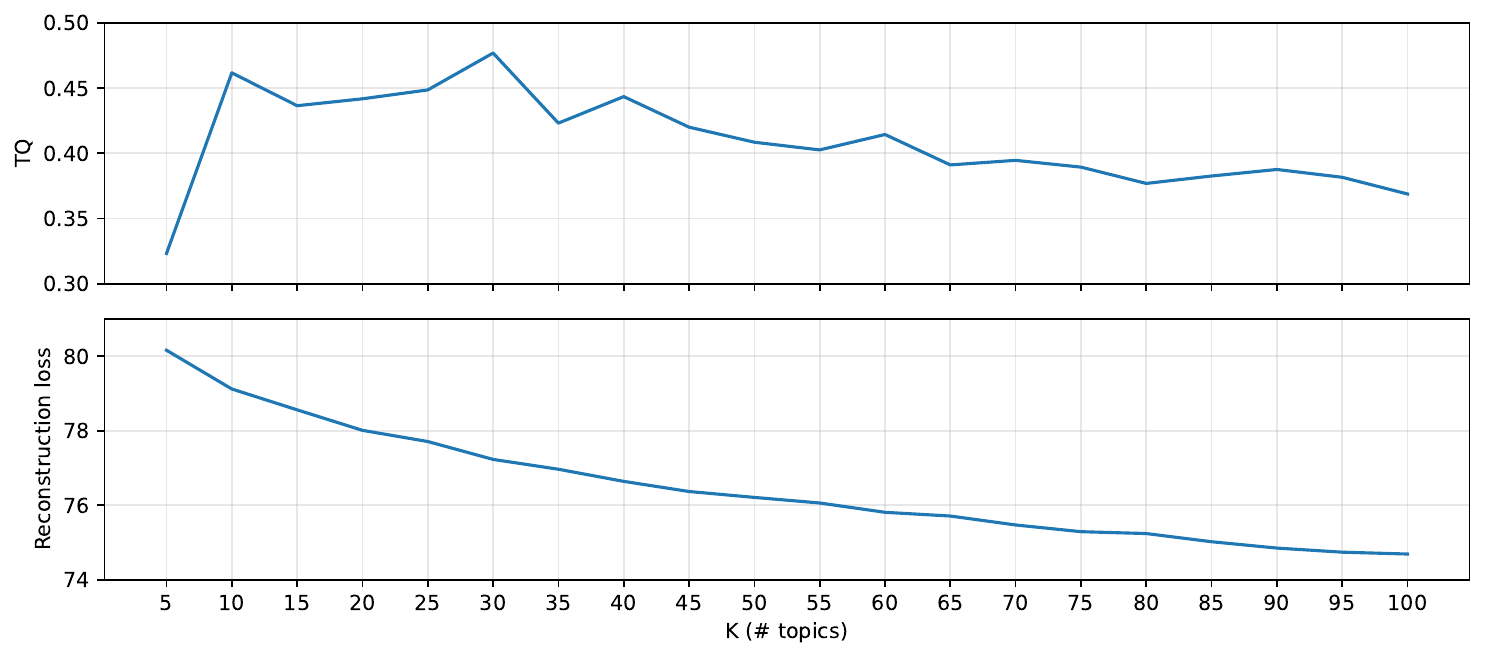}
    \caption{TQ \& reconstruction loss vs. number of topics ($K$).}
    \label{fig:choose_k}
\end{figure}

Figure~\ref{fig:choose_k} shows the trade-off between these metrics. As $K$ increases, reconstruction loss decreases steadily, indicating better data fitting. However, topic quality peaks at lower values of $K$ and declines as more topics are introduced, as topics become fragmented and less coherent. From a qualitative perspective, topics at moderate values of $K$ are more interpretable and distinct. Balancing these factors, we choose $K=60$ as the optimal value, representing a point where reconstruction loss begins to plateau while topic quality remains relatively high before dropping further.

\noindent\textbf{Post-Processing and Refinement:} The initial FASTopic output contains many highly granular, overlapping topics that hinder interpretation. To improve interpretability, we use Claude Sonnet 4.5 to label, merge semantically similar, and filter incoherent topics. This reduced topics from 60 to 37. Refinement details are in Appendix C, and the complete topic list with representative keywords is in Appendix D.

\section{RQ1: Topical Analysis}

We manually group topics into 3 categories: \textit{Grievance}, \textit{Policy}, and \textit{Other}, based on representative keywords and posts.

\noindent\textbf{Grievance:} encompasses rhetorical strategies to express political dissatisfaction, including complaints, criticism and negative remarks. \textit{Left-Wing Criticism} frames opponents through the liberal-conservative divide, attacking ``leftists,'' ``libs,'' and ``radicals,'' complemented by \textit{Critique of Authoritarian Ideologies}, which deploys labels like ``communist,'' ``socialist.''  \textit{Biden Criticism} features profane denunciations linking Biden to Obama, Hunter Biden, and Kamala Harris, characterizing the administration as the ``worst.'' \textit{Elections Integrity} centers on mobilization and voting participation, with calls to ``vote early,'' concerns about mail-in voting and ballot fraud. \textit{Primary Elections and Campaigns} highlights competitive races with discussion of candidates and campaign dynamics including figures like Dr.\ Oz. \textit{FBI and DOJ Investigations} centers on the Durham investigation, Hillary Clinton, and Hunter Biden's laptop, framing federal agencies as weaponized against Trump. \textit{January 6th Committee} dismisses the hearings as illegitimate, accuses the committee of altering evidence, and frames impeachment efforts as political theater. \textit{Supreme Court Decisions} debates Roe v.\ Wade, SCOTUS justices, and major rulings. \textit{Vulgar Political Insults} deploys profane language and mockery, targeting figures like AOC as ``clowns.'' \textit{Congress and Legislators} targets figures like Pelosi, McConnell, and congressional leadership. \textit{RINO and Establishment Critics} attacks perceived party traitors including Liz Cheney, Romney, and Bush as ``swamp'' figures. \textit{Cancel Culture and Media} addresses Hollywood, CNN, MSNBC boycotts, and entertainment industry politics. \textit{Mental Health Stigma} dismisses opponents as ``sick,'' ``evil,'' and hypocritical. \textit{Crime and Justice} demands accountability through arrests, charges, and imprisonment.

\noindent\textbf{Policy:} discussions that are shaped by moral and cultural narratives. \textit{Abortion} encompasses reproductive rights debates with references to women, birth, and life. A central theme is the Supreme Court's role in establishing and overturning Roe v. Wade. \textit{Education and Parental Rights} addresses anxieties about ideological influence on children through schools, with Disney as a focal point. \textit{Gun Policy} intertwines Second Amendment rights with discussions of political violence, mass shootings, and police while framing firearms as protection against government overreach. Discussion volume correlates with events such as Robb Elementary School shooting in late May 2022. \textit{Taxes and Government Spending} centers on frustrations over federal spending, IRS enforcement, and figures like Soros, while \textit{Economic Policy and Inflation} addresses rising costs, interest rates, and student loan debates. \textit{Energy Crisis Policy} ties fuel prices to Biden administration decisions on pipelines and oil production. \textit{COVID-19 and Vaccines} features skepticism toward Fauci, the CDC, and pandemic policies, linking vaccine concerns to broader medical establishment distrust. \textit{Border and Immigration} addresses illegal immigration, the southern border wall, and migrant policies.

\noindent\textbf{Others:} aggregates diverse themes from foreign policy to media consumption. \textit{Russia-Ukraine War} expresses skepticism toward foreign entanglements, questioning Ukraine funding, NATO expansion, and sanctions while debating Putin's motivations. \textit{US International Relations} discusses American positioning relative to major powers across Europe, the UK, and Asia. \textit{Government Power and Rights} addresses constitutional interpretation, federal overreach, and states' rights. \textit{States and Governors} contrasts red and blue state governance, elevating DeSantis and Florida while criticizing California. \textit{Business and Companies Issues} discusses corporate behavior, stock markets, and buying/selling dynamics. \textit{Political Party Conflict} examines Democrat-Republican dynamics and intra-party tensions. \textit{American Nationalism and Patriotism} links political identity with national pride, framing political struggle as defending America. \textit{Media Appearances and Events} promotes content on Fox News, live streams, and interviews, reinforcing collective identity through shared media consumption. \textit{Big Tech Censorship and Free Speech} reflects anxieties about platform censorship, celebrating Elon Musk's Twitter acquisition while discussing bans on Reddit and other platforms. \textit{Media Distrust and Misinformation} accuses mainstream media of ``fake news'' and propaganda, framing MSM narratives as deliberately deceptive. \textit{News and Press Releases} captures general news sharing and official announcements. \textit{Presidential Approval Ratings} compares Biden unfavorably to historical presidents, often invoking dementia narratives. \textit{Trump Support and Presidency} expresses devotion to Trump and MAGA, with mentions of DeSantis as a potential successor. \textit{Racial Politics Discussion} frames racial issues through accusations of racism and critiques of BLM. \textit{LGBTQ and Religious Issues} addresses transgender debates, gay marriage, and Christian religious liberty.

\subsection{Cross-Platform Comparison of Topics}

\begin{figure}[ht!]
\centering
\includegraphics[width=0.95\columnwidth]{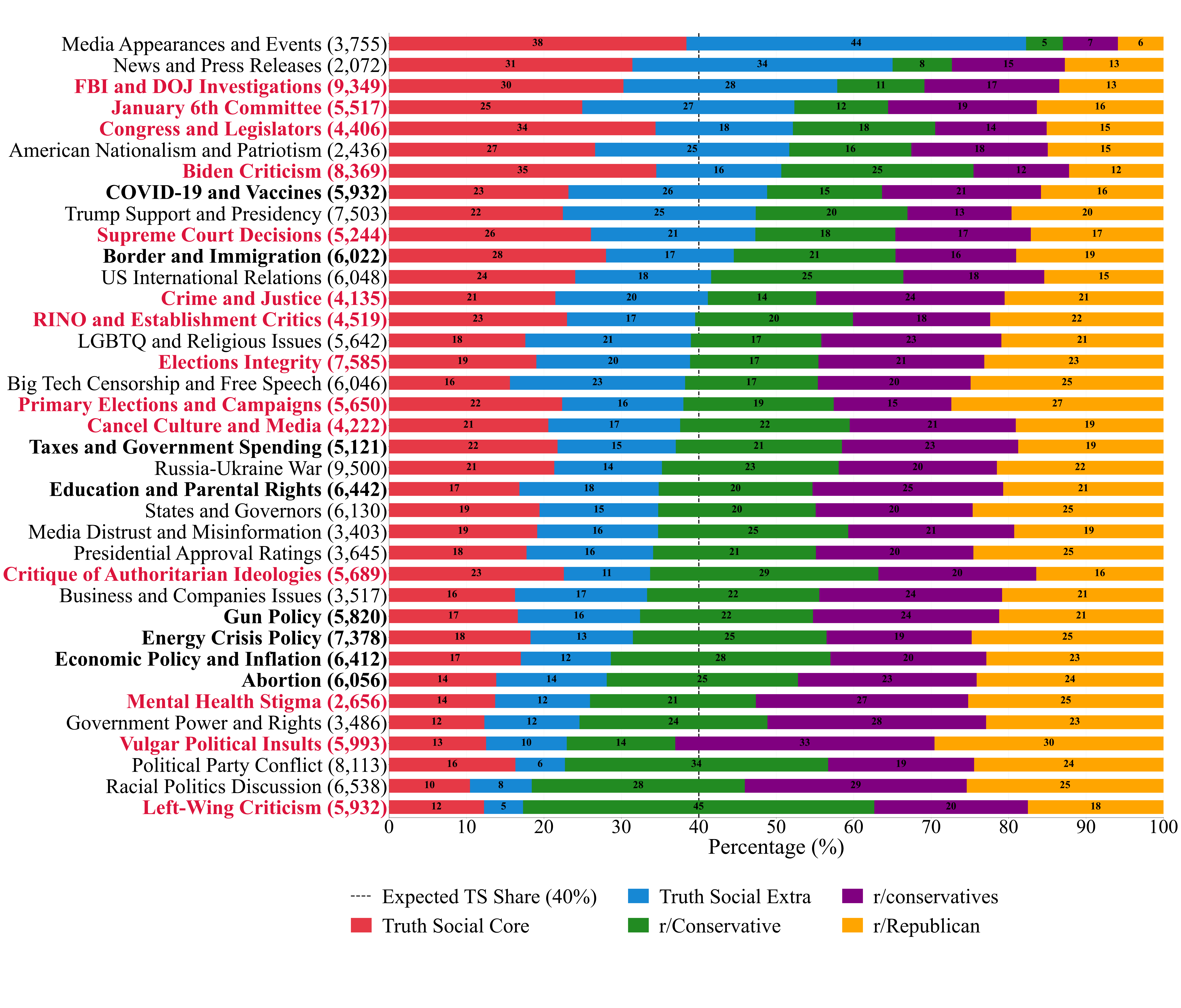}
\caption{Topic share by platform relative to the expected Truth Social share (40\%, vertical line). Label styles denote category: Red (Grievance), Bold (Policy), and Plain (Other).}

\label{fig:topic_source}
\end{figure}

For each topic, we compute the share of posts from each platform. Since we sample an equal number of posts from each platform, each platform is expected to account for 20\% of posts per topic. Figure~\ref{fig:topic_source} shows the results.

\noindent\textbf{Subreddit Differences:} The three subreddits show high alignment. In 20 of 37 topics, subreddits form a cohesive bloc where the difference between any pair is no more than 6\%. This alignment occurs primarily on consensus policy topics (e.g., \textit{Gun Policy}, \textit{Abortion}, \textit{Taxes and Government Spending}) and reactions to external threats (e.g., \textit{Biden Criticism}, \textit{Supreme Court Decisions}, \textit{Cancel Culture and Media}), where conservative consensus is unified and reactions are synchronized by the broader media ecosystem.

Substantial divergences emerge in categories related to moderation intensity and community function. r/Conservative stands as the primary outlier, diverging by more than 6\% from its counterparts on 10 topics. In contrast, r/conservatives and r/Republican display tighter correlation, with only 3 topics exceeding 6\% divergence.

The divergence on specific topics appears to stem from differing moderation strategies and discursive roles necessitated by community size and visibility. As the largest and most visible platform, r/Conservative faces greater external scrutiny, likely resulting in stricter civility enforcement. This is evident in the distribution of \textit{Vulgar Political Insults}, which constitute 33.47\% of discourse on r/conservatives and 29.57\% on r/Republican, but only 14.02\% on r/Conservative. Consequently, hostility toward the opposition on the flagship subreddit is channeled into the more sanitized \textit{Left-Wing Criticism}, capturing 45.35\% of conversation on r/Conservative, significantly higher than r/conservatives (19.84\%) or r/Republican (17.52\%). This inverse relationship suggests systemic critique serves as a permissible outlet for aggressive sentiment within the stricter moderation environment. This dynamic extends to other prominent ideological topics where r/Conservative leads, such as \textit{Biden Criticism} (the 3rd highest volume topic) and \textit{Critique of Authoritarian Ideologies}. This pattern may be partly attributable to the subreddit's ``flair'' system, which requires users to demonstrate ideological alignment to participate in conservative-only discussions. Such a mechanism likely fosters a more ideologically homogeneous and engaged user base, oriented toward reinforcing ideological boundaries through sustained criticism of perceived opponents. Meanwhile, r/Republican functions more as ``party machinery,'' diverging by focusing on procedural and electoral mechanics, showing higher engagement on topics like \textit{Primary Elections and Campaigns}, \textit{States and Governors}, and \textit{Presidential Approval Ratings}.

\noindent\textbf{Truth Social Differences:} Similarly, the two Truth Social datasets align within 5 percentage points for 22 of 37 topics. However, given their distinct sampling methods—Truth Social Core (Trump's direct followers and early adopters) versus Truth Social Extra (broader users sampled via keywords like ``MAGA'')—the internal divergences are instructive. Truth Social Core exhibits significantly higher shares in topics related to direct elite politics and specific policy battles, functioning as a political ``vanguard.'' It dominates conversation on \textit{Biden Criticism} (+18\%), \textit{Congress and Legislators} (+17\%), and \textit{Border and Immigration} (+12\%), reflecting a user base synchronized with Trump's daily political grievances. In contrast, Truth Social Extra represents the wider MAGA cultural ecosystem. Instead of focusing on specific political enemies, this group shifts to structural and cultural grievances, showing higher or comparable engagement on rights-based issues such as \textit{Big Tech Censorship and Free Speech} (+7\%), \textit{LGBTQ and Religious Issues} (+3\%), and \textit{COVID-19 and Vaccines} (+3\%). This suggests the broader user base is less tethered to minute-by-minute elite conflict driven by Trump's account and more motivated by systemic anxieties (e.g., censorship, ``woke'' culture). A key caveat is that greater diversity within Truth Social Extra generates many small or low-coherence topics, including those driven by single-hashtag posts (e.g., ``\#MAGA''). These topics are filtered out during quality control, resulting in Truth Social Core occupying a larger proportional share.

\noindent\textbf{Truth Social vs Reddit:} Truth Social resembles Twitter in design and affordances, functioning as a broadcast-oriented platform. It privileges one-to-many communication, making it conducive to news dissemination, political promotions, and elite signaling such as official statements and media appearances. Truth Social operates as a megaphone through which political elites communicate directly with followers, rather than a space for extended deliberation or peer debate. This broadcast orientation explains the platform's disproportionately high share of content in \textit{Media Appearances and Events} (41\% TS avg.\ vs.\ 6\% Reddit avg.) and \textit{News and Press Releases} (32\% TS avg.\ vs.\ 11\% Reddit avg.), reflecting its emphasis on top-down information flows.

This structure is intertwined with Truth Social's origins, launched by Donald Trump following his removal from mainstream social media platforms. Consequently, the platform disproportionately centers leader-driven narratives and grievance-based frames, reinforcing its function as a hub for Trump-aligned discourse. Topics related to perceived institutional persecution are especially prominent: \textit{FBI and DOJ Investigations} accounts for an average of 29\% of Truth Social discourse compared to 10\% on Reddit, while \textit{January 6th Committee} represents 26\% on Truth Social versus 16\% on Reddit. This overrepresentation underscores how Truth Social consolidates grievance narratives linked to Trump's political identity, amplifying them through a centralized and largely uncontested communication environment. Beyond explicitly Trump-aligned grievances, additional grievance-oriented topics including \textit{Congress and Legislators}, \textit{Biden Criticism}, \textit{Supreme Court Decisions}, and \textit{Crime and Justice} are also disproportionately salient on Truth Social, functioning as symbolic boundary markers that reinforce in-group cohesion through construction of external adversaries. Among topics where Truth Social exhibits dominance, 6 of 13 are grievance-oriented (46\%), while only 2 are policy-focused (15\%), showing the platform's orientation toward affective mobilization rather than policy deliberation.

By contrast, Reddit's threaded discussions, persistent comment hierarchies, and community-specific moderation facilitate extended deliberation, evidentiary support, and ideological contestation. This is evident in policy disputes where Reddit dominates: \textit{Abortion} (24\% Reddit avg.\ vs.\ 14\% TS avg.), \textit{Economic Policy and Inflation} (24\% vs.\ 15\%), \textit{Energy Crisis Policy} (23\% vs.\ 16\%), and \textit{Gun Policy} (22\% vs.\ 17\%) show higher concentration on Reddit, where discussion is framed around rights, moral reasoning, and policy consequences. Overall, 6 of 8 policy-related topics are dominated by Reddit, accounting for 26\% of all 23 Reddit-dominated topics.

Additionally, Reddit-dominated topics are overwhelmingly centered on conflictual and identity-driven discourse, highlighting the functional distinction between Truth Social's top-down broadcast model and Reddit's more volatile, peer-to-peer environment. While Truth Social serves as a disciplined megaphone for elite messaging, Reddit functions as a discursive arena for confrontation—a dynamic associated with higher levels of toxic political expression, examined in the next section. This structural divergence is evident in the heavy concentration of hostile topics on Reddit: \textit{Racial Politics Discussion} (27\% Reddit avg.\ vs.\ 9\% TS avg.), \textit{Vulgar Political Insults} (25\% vs.\ 11\%), and \textit{Left-Wing Criticism} (27\% vs.\ 13\%). These results suggest that while Truth Social's vertical structure amplifies leadership grievances, Reddit's horizontal structure incubates the movement's vitriol, serving as the primary site for cultural boundary-drawing and ideological policing.

A subset of topics remains evenly distributed across platforms, reflecting shared conservative concerns that cut across community boundaries. \textit{Border and Immigration} (22\% TS avg.\ / 21\% Reddit avg.), \textit{Election Integrity} (19\% / 19\%), and \textit{Russia--Ukraine War} (17\% / 19\%) exhibit balanced attention, with no single platform exercising dominance. These cases indicate that issues tied to national sovereignty, electoral legitimacy, and foreign policy resonate broadly across conservative ecosystems, transcending structural differences between platforms.

Overall, these findings illustrate how platform affordances and communicative norms produce distinct conservative discourse environments, with Truth Social privileging elite-centered grievance mobilization and Reddit supporting policy deliberation and conflictual identity work.

\subsection{Temporal Dynamics}
\begin{figure*}[!t]
    \centering

    \begin{subfigure}[b]{0.48\textwidth}
        \centering
        \includegraphics[width=\textwidth]{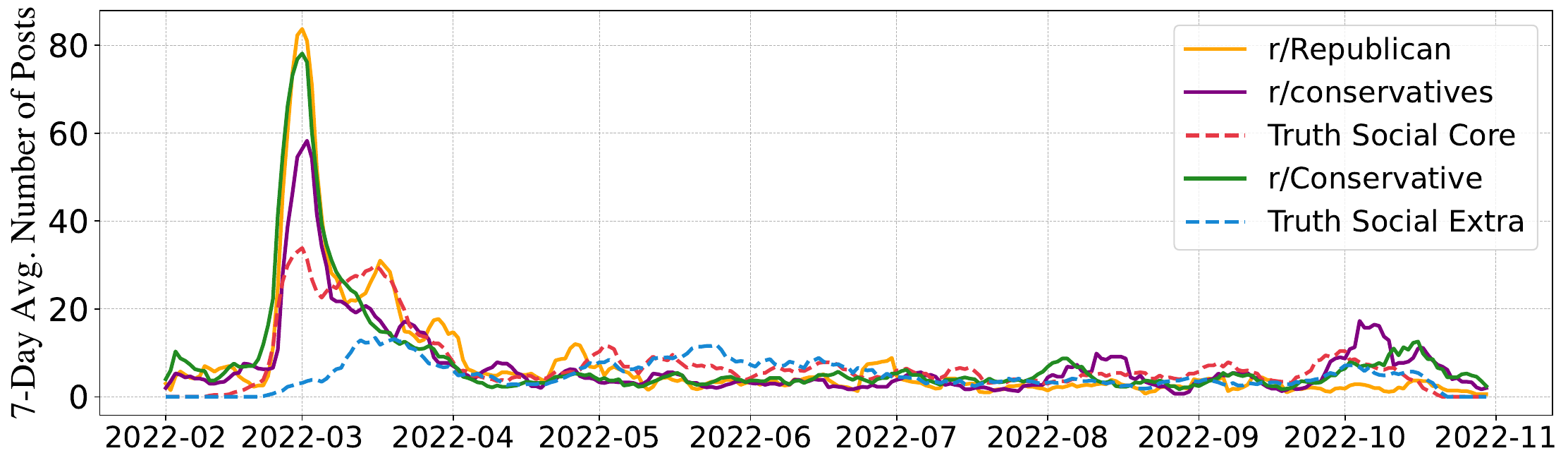}
        \caption{Russia-Ukraine War}
        \label{fig:sub_russia_ukraine}
    \end{subfigure}
    \hfill
    \begin{subfigure}[b]{0.48\textwidth}
        \centering
        \includegraphics[width=\textwidth]{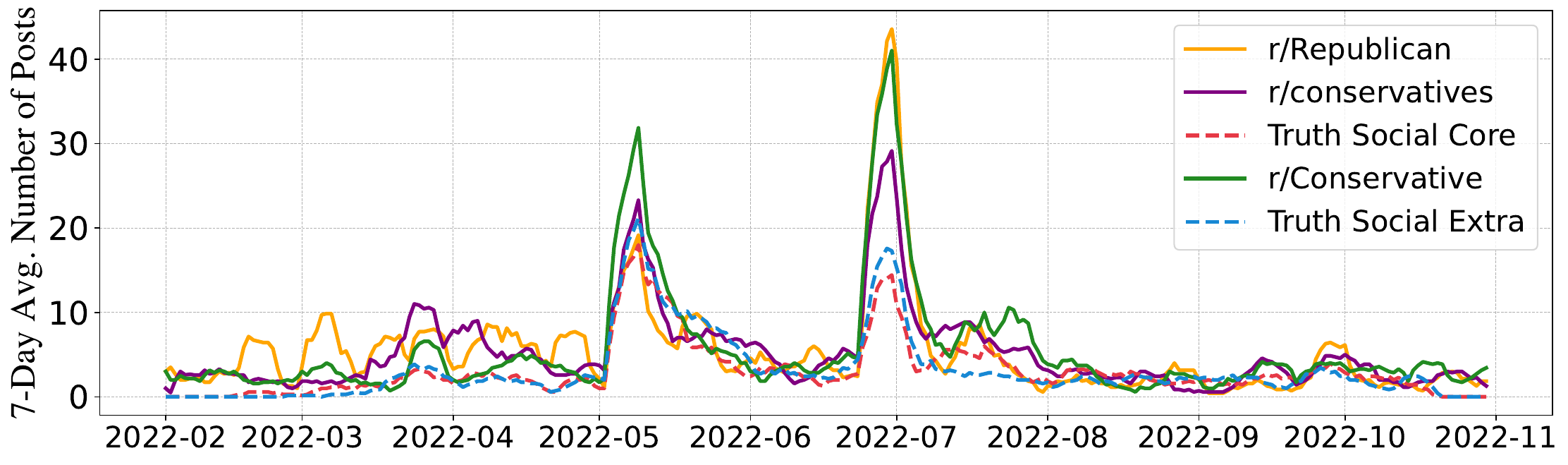}
        \caption{Abortion}
        \label{fig:sub_abortion}
    \end{subfigure}

    \vspace{0.15em}

    \begin{subfigure}[b]{0.48\textwidth}
        \centering
        \includegraphics[width=\textwidth]{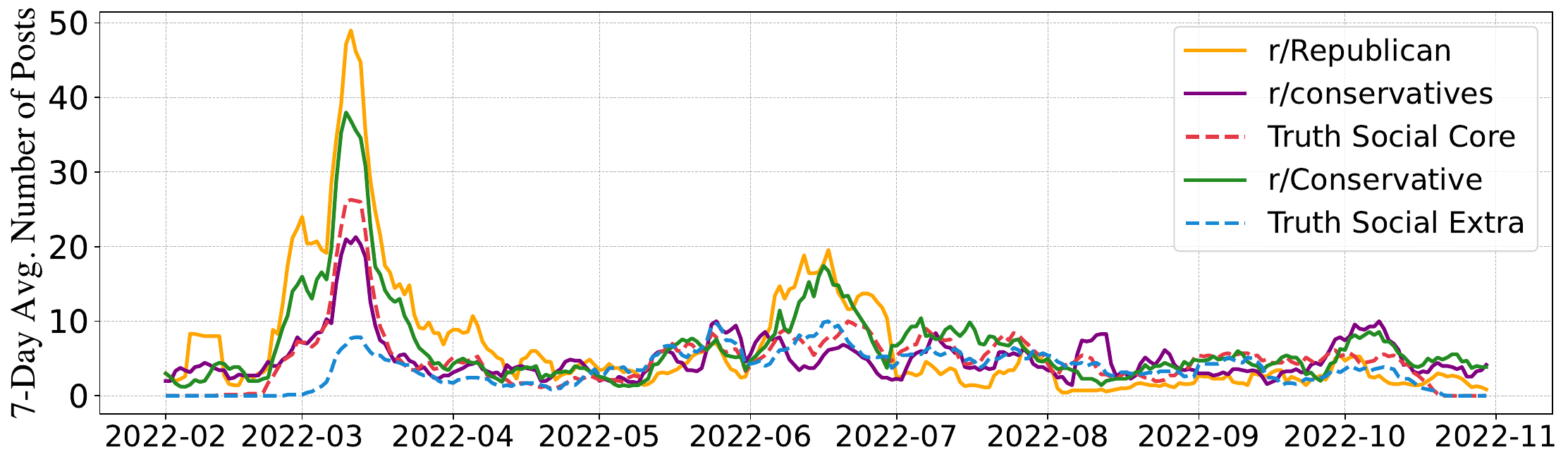}
        \caption{Energy Crisis Policy}
        \label{fig:sub_energy}
    \end{subfigure}
    \hfill
    \begin{subfigure}[b]{0.48\textwidth}
        \centering
        \includegraphics[width=\textwidth]{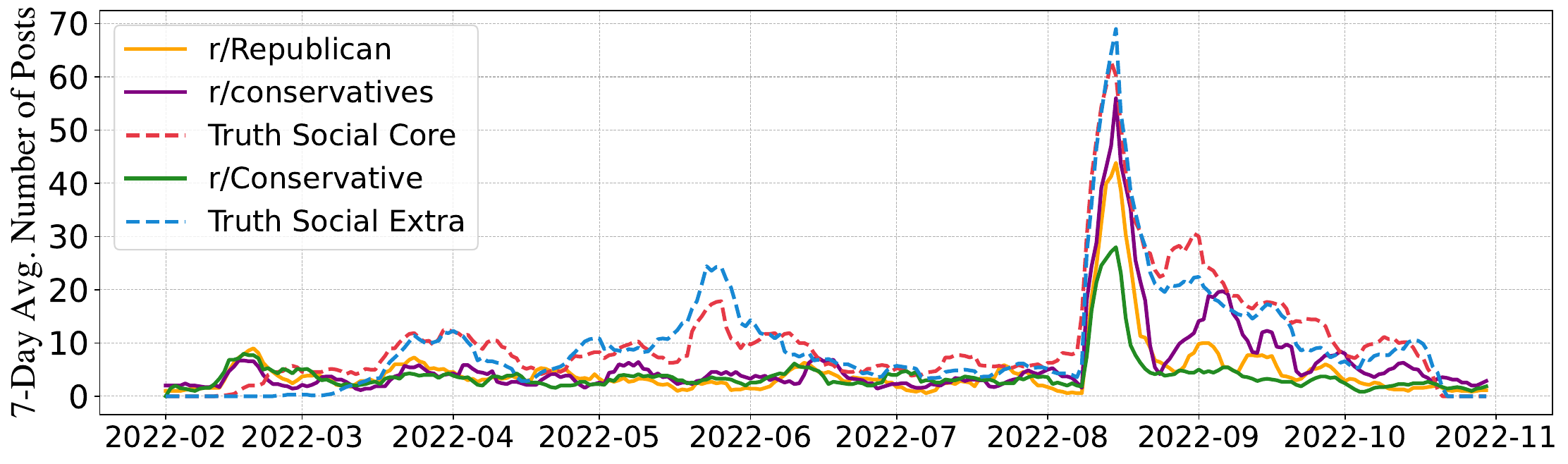}
        \caption{FBI and DOJ Investigation}
        \label{fig:sub_fbi}
    \end{subfigure}

    \caption{Temporal dynamics of document volume across four topics.}
    \label{fig:topic_dynamics_all}
\end{figure*}

We study how users discuss topics over time, identifying 8 topics that exhibit pronounced spikes around political events. Figure \ref{fig:topic_dynamics_all} shows the 7-day moving average number of posts per day for these topics. The remaining 4 are in Appendix E. To simplify the comparison, we merge the two Truth Social datasets in the remainder of the analysis.

We first investigate if these spikes are driven by users posting more frequently about a topic around event dates. We fit 8 Bayesian hierarchical time-series regression models with geometric likelihood and individual-level random intercepts to predict users' post counts across time and between platforms, whilst accounting for individual user differences. The analyses do not reveal group differences between platform user bases in terms of posting rates. Additionally, we do not observe an association of time or a change around event dates and the rates at which users posted about the topics. Thus, there is no evidence that the spikes are attributable to users posting more frequently around events, relative to their normal baselines. Results tables are in Appendix F.

Since changes in posting frequency do not explain the observed temporal patterns, we next test whether spikes are driven by increases in the proportion of users participating in a topic around major political events. We fit 8 Bayesian hierarchical time-series regression models with geometric likelihood and group-size offset, modeling changes in number of users posting about each topic over time across platforms. We center time around the date of the proposed event (measured in weeks, rounded to 2 decimal places) to estimate if the event caused a discontinuity in the proportion of users posting about each topic. Together, these analyses indicate that event-driven spikes are not driven by users posting more frequently, but rather an increase in the *proportion* of users participating in the topic.  Below, we report results for four topics. Interactions are tested with Leave-One-Out validation; we only report interactions when they improve out-of-sample model prediction. Results for Gun Policy, RINO and Establishment Critics, Border and Immigration, and Big Tech Censorship are reported in Appendix F, as they primarily reveal stable platform-level differences over time.

\noindent\textbf{Russia-Ukraine War:} The Russian invasion of Ukraine on February 24, 2022 produces a peak (Figure~\ref{fig:sub_russia_ukraine}). Controlling for group size, all subreddits exhibit lower proportions of users posting about the topic relative to Truth Social (r/conservative: RR=0.24, 95\% CrI[0.16–0.38], r/conservatives: RR=0.26, 95\% CrI[0.17–0.40], r/republican: RR=0.31, 95\% CrI[0.20–0.47]). Before the war began, we find a positive association between time and the proportion of users who posted about the topic (RR=1.20, 95\% CrI[1.01–1.42]). Specifically, for every 1 point increase in time there is an associated 20\% increase in the proportion of users posting about the topic. However, we caution that the lower bound of this estimate is very close to the null (1) and so it is likely the effect is so small it might be negligible. Moreover, at the time the war began there is a sudden 72\% increase in the proportion of users who posted about the Russia-Ukraine war (RR=1.72, 95\% CrI[1.09–2.77]) for all 4 platforms. These results suggest that while a higher proportion of Truth Social users discuss the war overall, the start of the war led to the same proportional increase for all 4 user bases.

\noindent\textbf{Abortion:} The U.S. Supreme Court's Dobbs decision overturning Roe v. Wade on June 24, 2022 coincides with a major surge in abortion-related discussion (Figure~\ref{fig:sub_abortion}). Controlling for group size, all subreddits show lower rates of users posting about abortion (r/conservative: RR=0.19, 95\% CrI[0.13–0.29], r/conservatives: RR=0.27, 95\% CrI[0.18–0.40], r/republican: RR=0.28, 95\% CrI[0.18–0.42]) relative to Truth Social. Although time is not associated with the proportion of users who posted about abortion prior to the overturning of Roe v. Wade (RR=0.96, 95\% CrI[0.82–1.14]), we find a sudden 119\% increase in the proportion of users who posted about abortion at the time Roe v. Wade was overturned (RR=2.19, 95\% CrI[1.34–3.57]). Thus, although abortion discussion is consistently more prevalent on Truth Social, the ruling itself generates a uniform participation shock across platforms.

\noindent\textbf{Energy Crisis Policy:} On March 8, 2022, the Biden Administration banned U.S. imports of Russian oil, gas and coal, a date which coincides with increased discussion (Figure~\ref{fig:sub_energy}). Controlling for group size, all 3 subreddits show lower proportions of users posting about energy crisis policy (r/conservative: RR=0.24, 95\% CrI[0.16–0.37], r/conservatives: RR=0.28, 95\% CrI[0.18–0.43], r/republican: RR=0.43, 95\% CrI[0.29–0.64]) relative to Truth Social. We find an interaction between time and the binary predictor (before/after the ban). Before the ban, for every 1 unit increase in time, there is a predicted 170\% increase in the proportion of users posting about the issue for all 4 platforms (RR=2.70, 95\% CrI[2.08–3.49]). At the time of the ban, there is a sudden 41\% decrease in the proportion of users posting about the topic. After the ban, every 1 unit increase in time predicts a 51.4\% decrease in the proportion of users posting about the energy crisis (RR=0.18, 95\% CrI[0.13–0.26]).

\noindent\textbf{FBI and DOJ Investigations:} On August 8, 2022, the FBI search of Trump's Mar-a-Lago estate triggered a major surge in discussion (Figure~\ref{fig:sub_fbi}). Controlling for group size, all subreddits show lower proportions of users posting about the topic (r/conservative: RR=0.07, 95\% CrI[0.05–0.10], r/conservatives: RR=0.16, 95\% CrI[0.11–0.24], r/republican: RR=0.13, 95\% CrI[0.08–0.19]) relative to Truth Social. We do not find an association between time and the proportion of users posting about FBI and DOJ investigations, except a steep 173\% increase at the time of the FBI search of Trump's Mar-a-Lago estate (RR=2.73, 95\% CrI[1.64–4.54]) for all 4 platform user bases.

\section{RQ2: Toxicity}

\begin{figure}[ht]
    \includegraphics[width=\columnwidth]{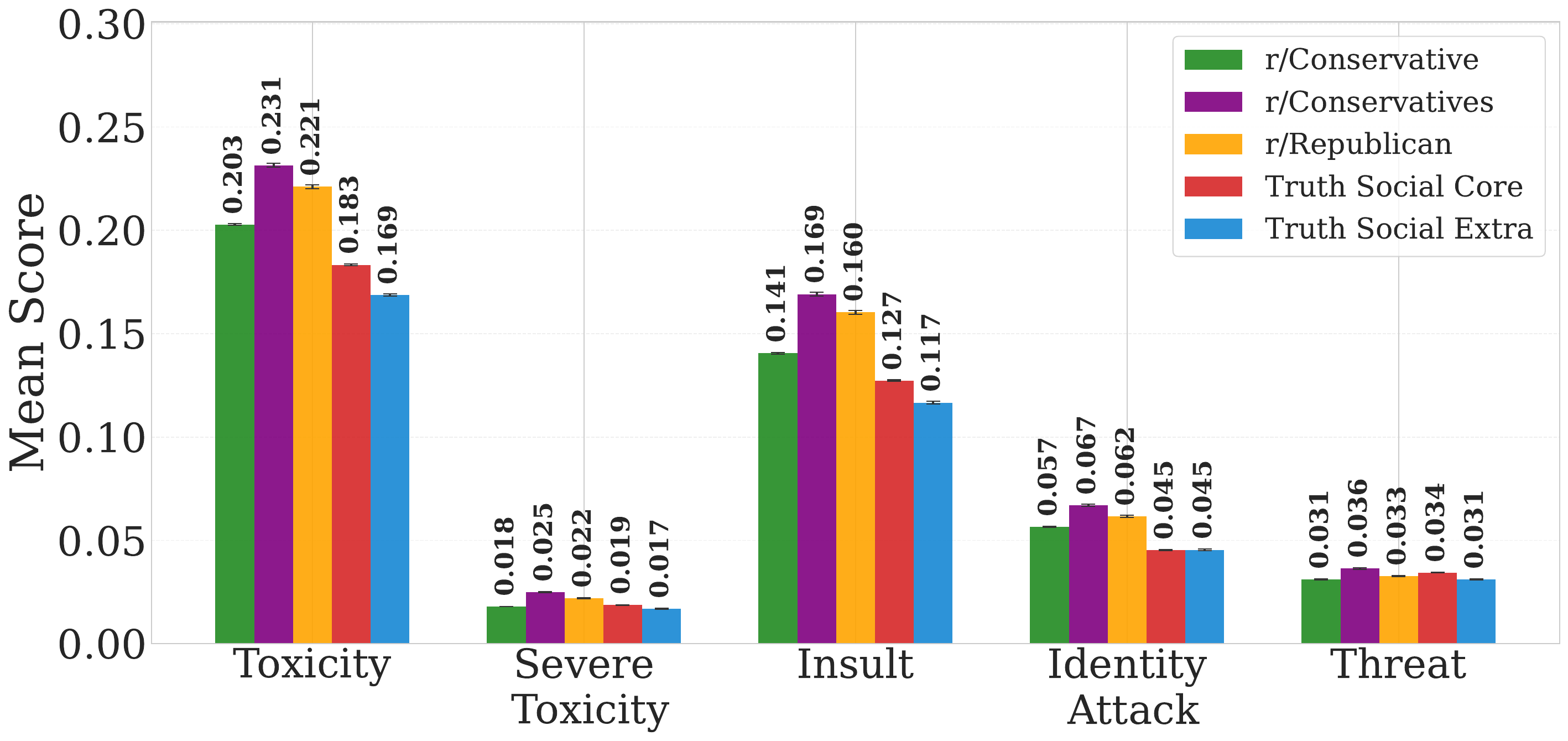}
    \caption{Mean toxicity scores across platforms.}
    \label{fig:toxicity_bars}
\end{figure}

We analyze toxicity levels across platforms and patterns across time. We first compute mean toxicity per platform by averaging the toxicity of posts. As Figure~\ref{fig:toxicity_bars} shows, the three subreddits exhibit slightly higher average toxicity than Truth Social datasets. This gap is most pronounced for overall toxicity and insults. Identity Attack, Severe Toxicity, and Threat are rather rare and platforms do not show high differences.

\begin{figure*}[ht!]
\centering
\includegraphics[width=0.9\textwidth]{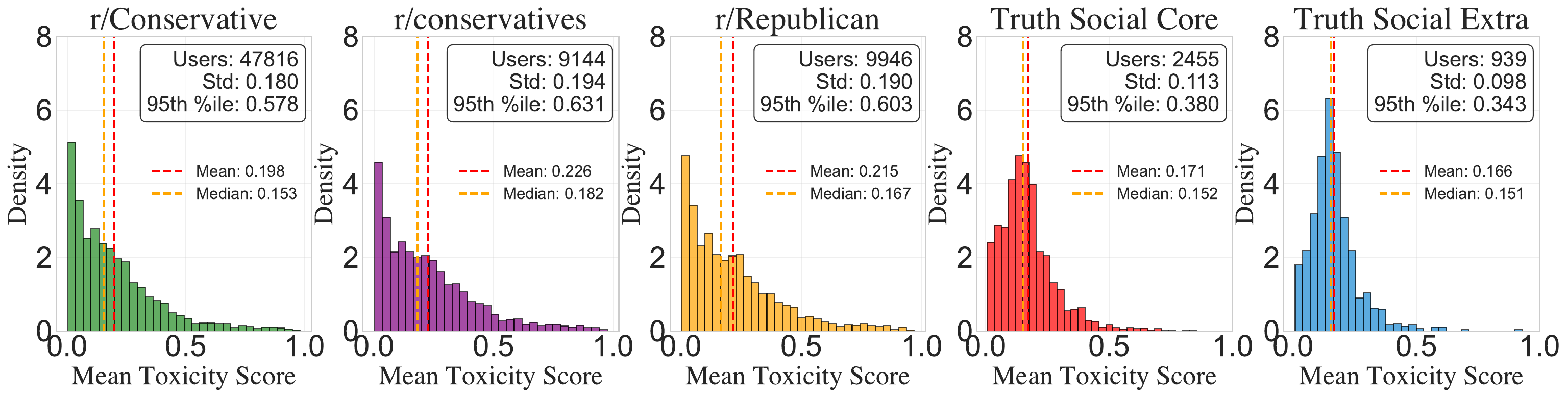}
\caption{Toxicity levels distribution per users per platform.}
\label{fig:tox_users}
\end{figure*}

We then compute each user's toxicity as the average of their post-level toxicity scores. The distributions across platforms are shown in Figure~\ref{fig:tox_users}. While both platforms show right-skewed distributions where the majority of users exhibit low toxicity, Truth Social's user base is considerably more homogeneous. Its users are tightly clustered at the low end of the toxicity scale, with a low standard deviation of 0.113 and a 95th percentile score of just 0.380 for \emph{Truth Social Core} dataset.
In contrast, Reddit displays greater heterogeneity in user toxicity, exhibiting much wider variance with higher standard deviations (e.g., 0.194 for r/conservatives) and a greater presence of high-toxicity outliers, pushing the 95th percentile to scores as high as 0.631.
Despite these differences in toxicity distributions, the platforms are similar in that high user activity does not correlate with high toxicity. Analysis of posting frequency on both platforms shows no strong positive correlation between the number of posts a user makes and their mean toxicity score.

The Bayesian Logistic Regression shows that, after accounting for user-level random effects, platform predicts the likelihood of posting toxic content. Models fit with only Truth Social and with both Truth Social datasets produce identical patterns, so results are reported using the combined Truth Social data. For all subreddit, the odds of users making a toxic post increase relative to Truth Social: 65\% for r/Conservative (OR=1.62, 95\% CrI[1.36–1.94]), 106\% for r/conservatives (OR=2.06, 95\% CrI[1.73–2.46]), and 98\% for r/republican (OR=1.98, 95\% CrI[1.67–2.34]). This shows Reddit is consistently more likely to be toxic than Truth Social, even after accounting for individual user differences.

\begin{figure}[ht!]
\centering
\includegraphics[width=0.96\columnwidth]{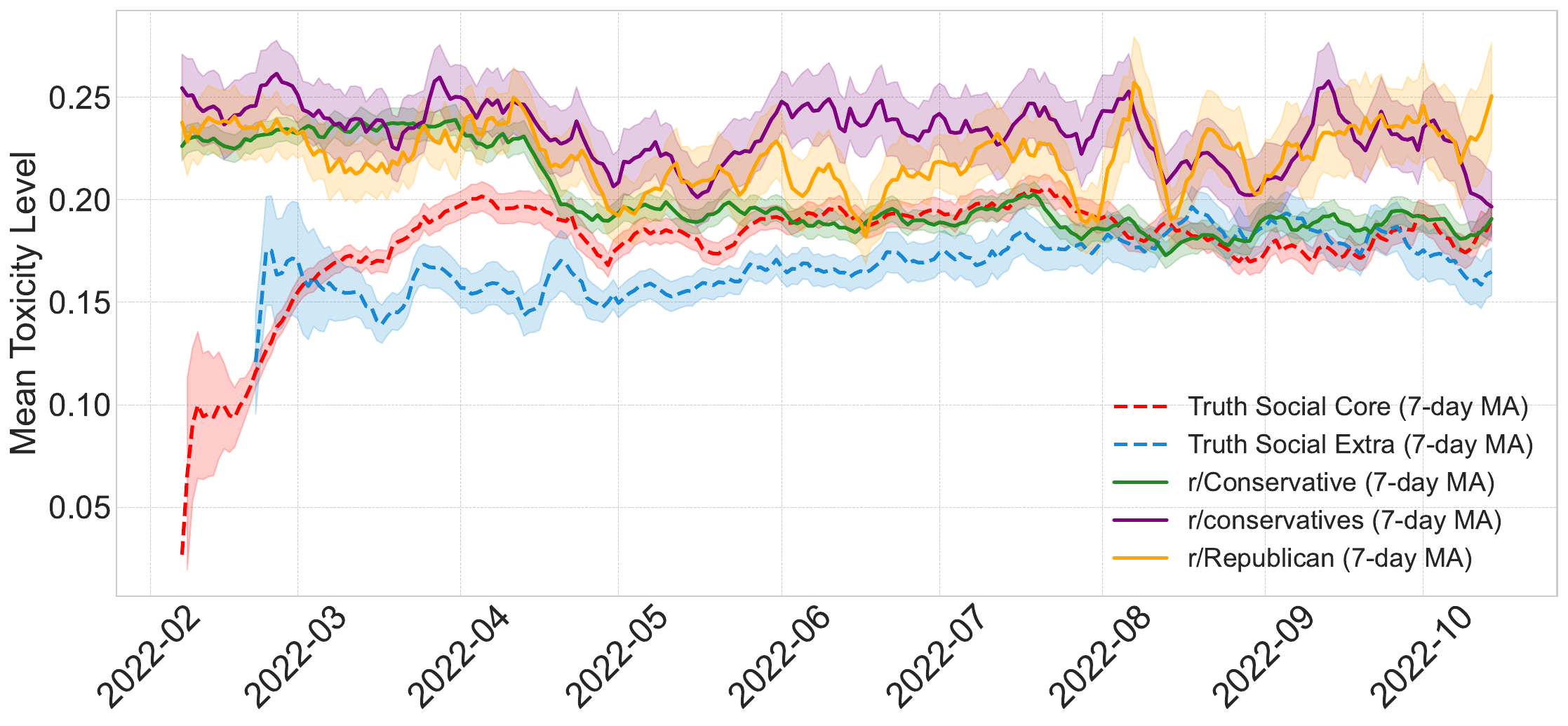}
\caption{Mean toxicity over time by platform. Shaded regions indicate $\pm 1$ standard error of the mean (SEM).}
\label{fig:temporal_toxicity}
\end{figure}

\noindent\textbf{Toxicity Temporal Dynamics:} Figure~\ref{fig:temporal_toxicity} illustrates temporal dynamics of mean toxicity levels across platforms. \emph{Truth Social Core} exhibited a dramatic escalation in mean toxicity from approximately 0.03 to 0.20 during February-March 2022, representing nearly a seven-fold increase. Similarly, \emph{Truth Social Extra} showed an increase in toxicity over the studied period, though with consistently lower levels compared to \emph{Truth Social Core}. r/Conservative demonstrates a decline from initial levels of approximately 0.23 to stabilize around 0.18-0.20, while r/conservatives maintains consistently elevated toxicity levels (0.24-0.26) throughout the observation period, and r/Republican shows high volatility with periodic spikes reaching 0.25.

\begin{figure}[ht!]
\centering
\includegraphics[width=0.85\columnwidth]{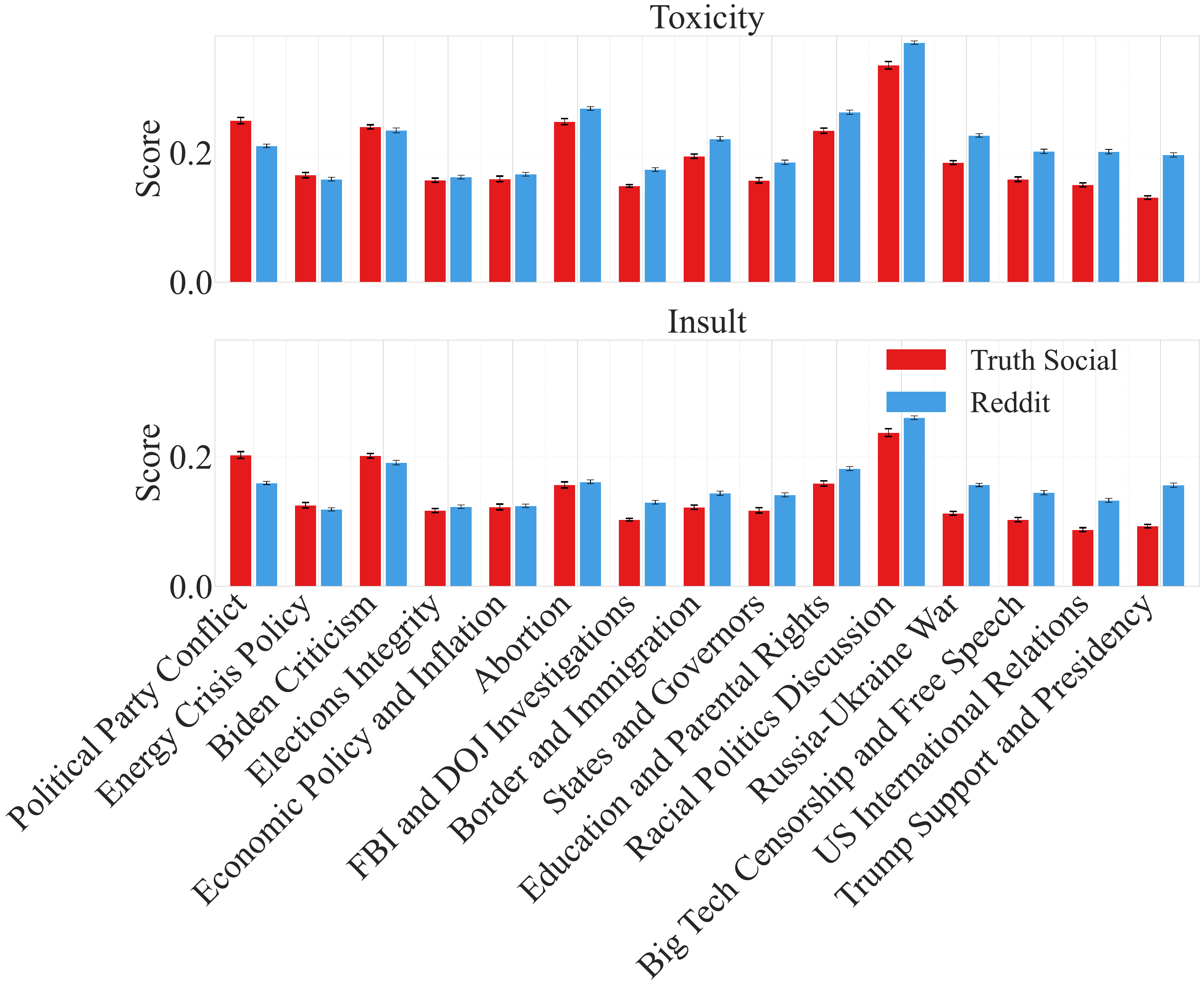}
\caption{Toxicity dimensions across the 15 largest topics.}
\label{fig:toxicity_topics}
\end{figure}

\noindent\textbf{Topics vs. Toxicity:} To study toxicity patterns between Truth Social and Reddit more closely, we analyze the 15 largest political topics, computing mean toxicity per topic for each platform. We merge both subreddits and Truth Social datasets, as there are no notable differences among them, and only present Toxicity and Insult as they are the most common types. Figure~\ref{fig:toxicity_topics} shows the results; while previous analysis established that Truth Social has lower mean toxicity, this does not hold for all topics.

Truth Social exhibits higher toxicity and insult scores than Reddit in a subset of ideologically charged domestic topics, particularly those centered on partisan conflict and critiques of political institutions. \textit{Political Party Conflict} shows higher toxicity on Truth Social (mean toxicity $=0.25$, insult $=0.20$) compared to Reddit (toxicity $=0.21$, insult $=0.16$). Similar patterns are observed in \textit{Biden Criticism} and \textit{Energy Crisis Policy}, where Truth Social marginally exceeds Reddit in both toxicity and insult.

In contrast, Reddit consistently exhibits higher toxicity and insult levels across the majority of remaining topics, particularly those related to social issues, governance, and international affairs. Discussions on \textit{Abortion}, \textit{Border and Immigration}, and \textit{FBI and DOJ Investigations} are more toxic on Reddit, with differences especially pronounced in \textit{Education and Parental Rights} (toxicity $=0.26$ on Reddit versus $=0.23$ on Truth Social). The most substantial disparity appears in \textit{Racial Politics Discussion}, where Reddit records the highest toxicity and insult scores overall (toxicity $=0.37$, insult $=0.26$), exceeding Truth Social (toxicity $=0.33$, insult $=0.24$). Similarly, Reddit shows markedly higher toxicity in discussions concerning the \textit{Russia--Ukraine War} and \textit{US International Relations}.

\section{Conclusion and Discussion}

This study provides a comparative analysis of conservative discourse on Truth Social and selected Reddit communities, focusing on topic, toxicity, and temporal dynamics. By systematically comparing Truth Social with conservative Reddit communities for the first time to the best of our knowledge, we contribute to prior research on echo chambers and online polarization. The topic modeling analysis shows that Truth Social discourse disproportionately focuses on leader-centered narratives, grievance politics, and symbolic themes, consistent with its role as a hub for mobilization and loyalty building. In contrast, conservatives on Reddit devote more space to policy debates, cultural controversies, and social identity issues, providing arenas for extended argument.

From the standpoint of political communication theory, the observed divergence between Truth Social and conservative Reddit communities reflects not only topical preferences but also the structuring role of platform affordances and governance in shaping political discourse. Elite-centered affordances are expected to emphasize grievance-oriented and personalized political narratives \cite{gillespie2018custodians, van2013understanding}. In contrast, platforms organized around threaded discussions and community moderation are more likely to foreground policy debates, ideological contestation, and cultural conflict, as discourse architectures and governance arrangements structure how political disagreement is expressed \cite{freelon2015discourse, matias2019civic, lampe2004slash}. Beyond its empirical contributions, this study advances political communication theory by demonstrating how platform affordances and moderation regimes shape not only the intensity of political discourse, but its dominant communicative form. By empirically distinguishing grievance-centered narrative spaces from policy-oriented deliberative arenas, we refine existing accounts of polarization, echo chambers, and alternative platform dynamics. Our findings show that alternative platforms such as Truth Social do not merely amplify toxic discourse, but restructure political communication around sustained grievance narratives and elite-centered engagement.

\noindent\textbf{Limitations:} The dataset covers a specific time frame (February–October 2022) and may not capture longer-term shifts in discourse. The analysis focuses on a subset of platforms and does not include other important conservative social media spaces such as Facebook, X, or YouTube. Finally, while the study identifies structural patterns and associations of user activity across platforms, it may not be representative of all the users on the platforms (e.g., less active or non-posting users) or conservatives in general.

\bibliography{aaai25}

\section{Checklist}

\begin{enumerate}

\item For most authors...
\begin{enumerate}
    \item Would answering this research question advance science without violating social contracts, such as violating privacy norms, perpetuating unfair profiling, exacerbating the socio-economic divide, or implying disrespect to societies or cultures?\\
    \answerYes{Yes, the study advances understanding of online political discourse using a preexisting ICWSM Truth Social data and public Reddit dataset. It does not interfere with the privacy of users or harm them in an another way.}

  \item Do your main claims in the abstract and introduction accurately reflect the paper's contributions and scope?\\
    \answerYes{Yes, the abstract and introduction summarize the analysis of the methodology and findings related to topics, toxicity and temporal dynamics across the platforms.}

  \item Do you clarify how the proposed methodological approach is appropriate for the claims made?\\
    \answerYes{Yes, the methodological approach is explicitly linked to the claims made. The Data Labeling section explains how political content, leanings, and toxicity were labeled, ensuring reliable inputs. The topic modeling methodology outlines how topics were discovered, refined, and validated using quantitative metrics, with extensive details of LLM-assisted labeling in Appendix. Toxicity and platform modeling with respect to it also described in methodology. These ensure that the methods directly align with and justify the study's claims.}

  \item Do you clarify what are possible artifacts in the data used, given population-specific distributions?  \\
    \answerYes{Yes, we acknowledge that Truth Social and Reddit communities attract different user bases, which may bias results. This limitation is discussed in Data and Limitations sections.}

  \item Did you describe the limitations of your work? \\
    \answerYes{Yes, in Conclusion.}

  \item Did you discuss any potential negative societal impacts of your work?  \\
    \answerYes{Yes, we note that analyzing toxicity may inadvertently attribute toxicity to users in a platform and clarify that this is not our purpose.}

  \item Did you discuss any potential misuse of your work?  \\
    \answerYes{Yes, we highlight that findings on toxicity could be misused to stereotype communities. We caution against simplistic generalizations and emphasize analytical context.}

  \item Did you describe steps taken to prevent or mitigate potential negative outcomes of the research, such as data and model documentation, data anonymization, responsible release, access control, and the reproducibility of findings?  \\
    \answerNA{
    We did not release or redistribute any datasets derived from the platforms studied. All data were processed in aggregate form, and no attempts were made to identify or profile individual users. To further ensure robustness, we report results across multiple random seeds and focus on comparative and temporal patterns rather than individual-level predictions.
    }

  \item Have you read the ethics review guidelines and ensured that your paper conforms to them?  \\
    \answerYes{Yes, the work follows ICWSM ethics guidelines by relying on public data, avoiding personal identifiers, and contextualizing results responsibly.}
\end{enumerate}

\item Additionally, if your study involves hypotheses testing...
\answerNA{NA ,  our study is exploratory and descriptive, focusing on comparative discourse analysis rather than formal hypothesis testing.}

\item Additionally, if you are including theoretical proofs...
\answerNA{NA ,  our study does not include theoretical results or proofs.}

\item Additionally, if you ran machine learning experiments...
\begin{enumerate}
  \item Did you include the code, data, and instructions needed to reproduce the main experimental results (either in the supplemental material or as a URL)? \\
    \answerNA{The dataset is public. We will release the code upon acceptance}

  \item Did you specify all the training details (e.g., data splits, hyperparameters, how they were chosen)?  \\
    \answerYes{Yes, Methodology describes model parameters, including Mallet LDA, Bertopic, FASTopic settings, and evaluation metrics.}

  \item Did you report error bars (e.g., with respect to the random seed after running experiments multiple times)?\\
    \answerYes{Yes, Table~\ref{tab:model_comparison} reports means with standard deviations across multiple runs.}

  \item Did you include the total amount of compute and the type of resources used (e.g., type of GPUs, internal cluster, or cloud provider)?  \\
    \answerYes{Yes, topic modeling and toxicity scoring were run on standard academic computing clusters.}

  \item Do you justify how the proposed evaluation is sufficient and appropriate to the claims made? \\
    \answerYes{Yes, we combine coherence, diversity, and topic quality metrics with qualitative inspection to ensure robust evaluation.}

  \item Do you discuss what is “the cost” of misclassification and fault (in)tolerance?  \\
    \answerYes{Yes, Limitations notes that toxicity scores may contain misclassifications; results are interpreted at aggregate rather than individual levels to mitigate harm.}
\end{enumerate}

\item Additionally, if you are using existing assets (e.g., code, data, models) or curating/releasing new assets...
\begin{enumerate}
  \item If your work uses existing assets, did you cite the creators? \\
    \answerYes{Yes, we cite the creators of all datasets, models, and tools used in the study.}

  \item Did you mention the license of the assets?  \\
    \answerYes{Yes, licenses are acknowledged where applicable (e.g., Perspective API terms, open-source toolkits).}

  \item Did you include any new assets in the supplemental material or as a URL?  \\
    \answerNA{NA,  we do not release new datasets due to platform restrictions, but provide processed outputs at the topic level.}

  \item Did you discuss whether and how consent was obtained from people whose data you're using/curating?  \\
    \answerYes{Yes, all data are from public datasets that are accessible online. Users post public voluntarily; no private data were collected.}

  \item Did you discuss whether the data you are using/curating contains personally identifiable information or offensive content? \\
\answerYes{Yes, we state that the data are publicly available posts without personally identifiable information, but they do contain offensive and toxic content, which we address in the analysis.}

  \item If you are curating or releasing new datasets, did you discuss how you intend to make your datasets FAIR?
    \answerNA{NA ,  no new dataset is released.}

  \item If you are curating or releasing new datasets, did you create a Datasheet for the Dataset?
    \answerNA{NA ,  no new dataset is released.}
\end{enumerate}

\item Additionally, if you used crowdsourcing or conducted research with human subjects...
\answerNA{NA ,  this study relies only on publicly available online data and does not involve direct human subjects research.}

\end{enumerate}

\section{Ethics Statement} We only analyze public profiles and present aggregate results. The toxicity classifier may misclassify, so findings should be read at the aggregate level, not as identity attacks or claims that all like-minded individuals are toxic.

\appendix

\section*{Appendix A: Platform Growth}
\addcontentsline{toc}{section}{Appendix A: Platform Growth}
\label{appendix:growth}

\begin{figure}[ht!]
\centering
\includegraphics[width=\columnwidth]{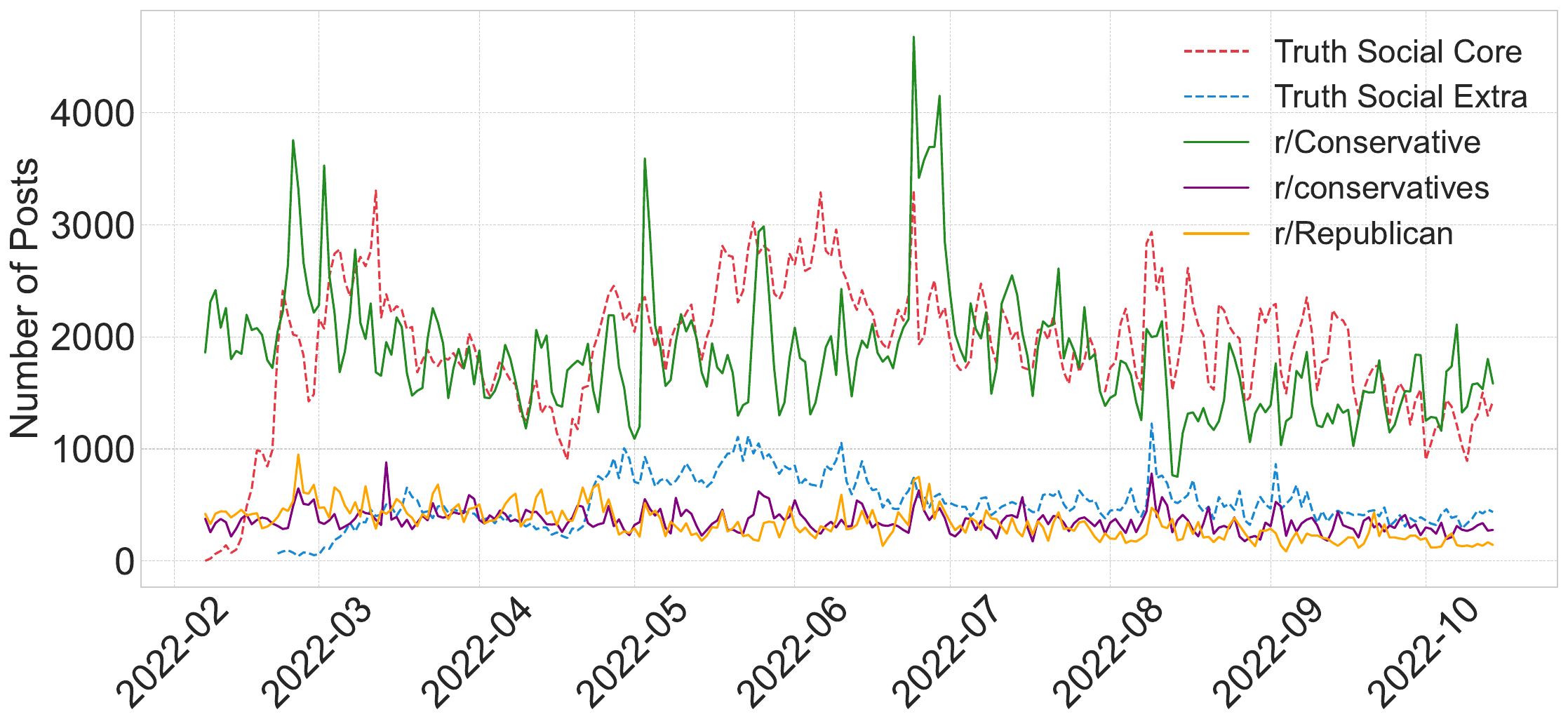}
\caption{Number of posts per day across platforms. }
\label{fig:daily_counts}
\end{figure}

\begin{figure}[ht!]
    \centering
    \includegraphics[width=0.96\columnwidth]{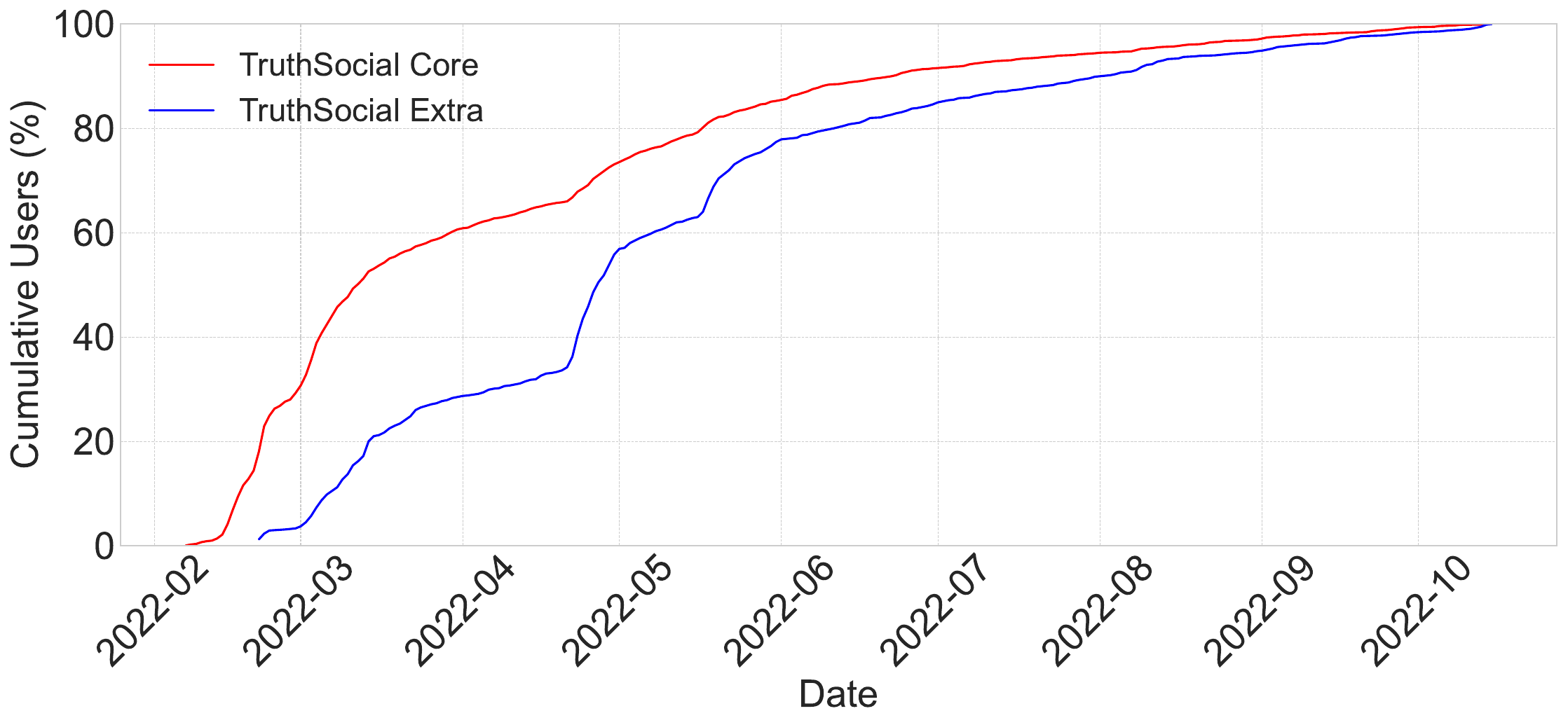}
    \caption{Cumulative percentage of new users over time for Truth Social Core and Truth Social Extra. The plateauing of the curves indicates the transition from rapid initial adoption to platform maturity by mid-2022.}
    \label{fig:daily_counts_cum}
\end{figure}

To address concerns regarding the temporal scope of this study, we analyze daily posting activity and user registration trends during the observation period. Registration is based on the date of the first post by a user since the datasets lacked information about the joining date of users. Although the dataset focuses on an eight-month window in 2022, Figure~\ref{fig:daily_counts} and Figure~\ref{fig:daily_counts_cum} indicates that Truth Social rapidly achieved platform maturity. Figure~\ref{fig:daily_counts} shows the daily activity of the five data sources between February and October 2022. Figure~\ref{fig:daily_counts} shows that r/Conservative and \emph{Truth Social Core} sustain much higher activity, typically between 1,000 and 1,500 posts per day. \emph{Truth Social Extra} follows a broadly similar temporal trajectory after April 2022, although at a consistently lower volume, reflecting its smaller user base.

Figure~\ref{fig:daily_counts_cum} shows that the cumulative distribution of new users exhibits a sharp initial increase followed by a clear inflection point in late Q2 2022. As the number of new joiners stabilized shortly after the launch surge, the platform entered a steady-state phase. This suggests that the period of analysis captures the establishment of the platform’s core functional and ideological user base. Consequently, the findings represent the stabilized environment of the platform rather than a transient launch phase.

\appendix
\section{Appendix B:Topic Modeling Implementation Details}
\label{app:topic_modeling}

\noindent\textbf{LDA:} As a probabilistic generative model, LDA infers latent themes by modeling documents as mixtures of topics and topics as distributions over words. We use a Gibbs-sampling implementation of latent Dirichlet allocation trained on a bag-of-words representation. The Dirichlet prior $\alpha$ for document–topic distributions is set to $1/K$, following the standard heuristic~\cite{blei2003latent}, where $K$ is the number of topics. We train the model for 2000 iterations with an optimization interval of 50 and use a fixed random seed for reproducibility.

\noindent\textbf{BERTopic:} Combines contextual embeddings with clustering to discover semantically coherent themes. We use the standard setup which combines transformer-based embeddings, dimensionality reduction, and clustering~\cite{grootendorst2022bertopic}. To improve interpretability, we enable the refinement option, which re-estimates representative words using KeyBERT-inspired relevance, Maximal Marginal Relevance (MMR), and part-of-speech filtering to balance the frequency, relevance, and linguistic plausibility.

\noindent\textbf{FASTopic:} Embeds documents, topics, and words in a shared semantic space to improve coherence and coverage. We use the original implementation which embeds documents, topics, and words in a shared semantic space~\cite{wu2024fastopic}. It employs Sinkhorn distances to approximate optimal transport between documents and topics, as well as topics and words~\cite{cuturi2013sinkhorn}. We tune the entropic regularization parameter $\alpha_{dt}$ to improve stability and ensure broad topic coverage in our experiments.

\section*{Appendix C: LLM-Assisted Topic Refinement}
\addcontentsline{toc}{section}{Appendix C: LLM-Assisted Topic Refinement}
\label{app:llm_prompt_c}

\paragraph{Scope.}
We describe how we reduced topics from 60 to 37 and provide the exact prompt used for labeling.

\subsection*{Step 1: LLM-Assisted Labeling}
We first apply LLM-assisted labeling with Claude Sonnet 4.5 accessed via API, to generate interpretable descriptions for the initial 60 topics.
For each topic, we submit its ten keywords and twenty most representative posts (ranked) through a structured prompt.
The API returns results in a standardized JSON schema with:
(i) a primary label (2–5 words),
(ii) up to three alternatives,
(iii) a confidence score (0–1), and
(iv) a short rationale with supporting terms and post indices.
This automated process produces consistent summaries while providing transparency through confidence scores.
The full prompt used for this stage is provided at the end of this appendix.

\subsection*{Step 2: Semantic Merging}
Next, we reduce repetition by merging topics that show strong semantic overlap.
We compare each topic against others using overlap in top keywords and representative posts and cross-validation with LLM-generated labels.
Beyond labeling, we also use the LLM to assess whether topics capture the same underlying theme, providing a second layer of judgment beyond statistical measures. Following this, we merge two clusters on the \textit{Vulgar Political Insults} into a single broader category
For the merged topic, we combine the keywords and representative documents from the original clusters so that the new topic preserves the full thematic coverage of its components.

\subsection*{Step 3: Filtering Incoherent or Underrepresented Topics}
Finally, we remove topics that are incoherent or too small for reliable analysis.
We identify 16 topics as ``Mixed/Multiple Themes'' with confidence scores below 0.3, indicating weak coherence.
These topics include unrelated keywords and posts without a clear theme.
We also exclude six topics with fewer than 700 posts, well below the median topic size of 2,641.
They likely reflect noise or overly narrow micro-themes.
We avoid forcing interpretation onto weak clusters and maintain statistical reliability in the final topic set by removing peripheral topics.

\addcontentsline{toc}{section}{Appendix C}

\begin{tcolorbox}[breakable, colback=gray!3, colframe=black!20, title=LLM Prompt]

\textbf{Task.} You are a topic labeler. Your task: give one short, clear label for a topic using the top terms and the top 20 documents as context.

\subsubsection*{Rules}
\begin{enumerate}[left=0pt,itemsep=2pt,topsep=4pt]
    \item \textbf{Length:} 2–5 words, Title Case (e.g., ``Electric Vehicle Policy'').
    \item \textbf{Specificity:} Use clear concepts, not single events, unless the topic is an event.
    \item \textbf{Avoid:} Dates, IDs, vague or generic labels.
    \item \textbf{Entities:} Use categories (e.g., ``Elections'') unless a single named entity is central.
    \item \textbf{Mixed/Noisy Topics:} Use ``Mixed / Multiple Themes'' and add one sentence of explanation.
    \item \textbf{Evidence:} Provide a confidence (0–1), a 1–2 sentence rationale, and list 2–4 supporting terms and 2–4 supporting doc indices.
\end{enumerate}

\subsubsection*{Output Format}
\noindent\textit{Return exactly this JSON:}

\lstdefinelanguage{json}{
    basicstyle=\ttfamily\small,
    numbers=none,
    showstringspaces=false,
    breaklines=true,
    literate=
     *{0}{{{\color{black}0}}}{1}
      {1}{{{\color{black}1}}}{1}
      {2}{{{\color{black}2}}}{1}
      {3}{{{\color{black}3}}}{1}
      {4}{{{\color{black}4}}}{1}
      {5}{{{\color{black}5}}}{1}
      {6}{{{\color{black}6}}}{1}
      {7}{{{\color{black}7}}}{1}
      {8}{{{\color{black}8}}}{1}
      {9}{{{\color{black}9}}}{1}
}
\lstset{
    language=json,
    frame=single,
    rulecolor=\color{black!20},
    columns=fullflexible,
    keepspaces=true
}
\begin{lstlisting}
{
  "topic_id": "[TOPIC_ID]",
  "label": "",
  "alt_labels": [],
  "confidence": 0.0,
  "rationale": "",
  "evidence": {
    "terms": [],
    "docs": []
  },
  "keywords": []
}
\end{lstlisting}

\subsubsection*{Notes}
\begin{itemize}[left=0pt,itemsep=2pt,topsep=2pt]
    \item \texttt{"alt\_labels"}: up to 3 alternatives (may be empty).
    \item \texttt{"keywords"}: 3–6 salient keywords for search/tagging.
    \item Do not include anything else outside the JSON.
\end{itemize}

\textbf{Inputs}
\begin{tabularx}{\textwidth}{@{}p{2.8cm}X@{}}
\textbf{Topic ID} & \texttt{[TOPIC\_ID]} \\
\textbf{Top Terms} & \texttt{[10 KEYWORDS]} \\
\textbf{Top 20 Docs} & \texttt{[20 DOCUMENT EXCERPTS]} \\
\end{tabularx}
\end{tcolorbox}

\addcontentsline{toc}{section}{Appendix D}

\begin{table}[H]
\centering
\caption*{\textbf{Appendix D: Topics and Representative Keywords}}

\footnotesize
\begin{tabular}{p{0.25\textwidth} p{0.70\textwidth}}
\toprule
\textbf{Topic} & \textbf{Representative Keywords} \\
\midrule
Abortion & woman, abortion, baby, men, life, birth, child, female, sex, gender \\
\addlinespace
American Nationalism and Patriotism & american, country, america, world, united, nation, destroy, destroying, democracy, usa \\
\addlinespace
Biden Criticism & biden, joe, president, obama, administration, hunter, kamala, harris, admin, worst \\
\addlinespace
Big Tech Censorship and Free Speech & twitter, post, social, free, elon, banned, musk, sub, comment, reddit \\
\addlinespace
Border and Immigration & border, illegal, immigrant, southern, illegals, open, immigration, migrant, alien, wall \\
\addlinespace
Business and Companies Issues & company, business, buy, food, big, buying, sell, corporation, stock, private \\
\addlinespace
Cancel Culture and Media & cnn, movie, cancel, hollywood, msnbc, tv, star, network, actor, boycott \\
\addlinespace
Congress and Legislators & house, pelosi, congress, senate, nancy, senator, rep., paul, congressional, mcconnell \\
\addlinespace
COVID-19 and Vaccines & vaccine, health, death, fauci, drug, medical, cdc, pandemic, doctor, dr. \\
\addlinespace
Crime and Justice & crime, criminal, charge, jail, prison, arrested, arrest, charged, held, defund \\
\addlinespace
Critique of Authoritarian Ideologies & communist, nazi, socialist, commie, marxist, fascist, communism, authoritarian, socialism, dictator \\
\addlinespace
Economic Policy and Inflation & inflation, policy, economy, pay, tax, loan, cost, government, rate, high \\
\addlinespace
Education and Parental Rights & school, child, kid, parent, teacher, disney, young, student, old, age \\
\addlinespace
Elections Integrity & vote, election, voter, voting, voted, ballot, fraud, poll, million, stolen \\
\addlinespace
Energy Crisis Policy & gas, oil, price, energy, climate, fuel, green, production, electric, pipeline \\
\addlinespace
FBI and DOJ Investigations & fbi, hillary, clinton, doj, document, raid, investigation, evidence, laptop, agent \\
\addlinespace
Government Power and Rights & government, law, state, federal, right, control, power, constitution, amendment, rule \\
\addlinespace
Gun Policy & gun, police, violence, shooting, protest, mass, violent, murder, terrorist, attack \\
\addlinespace
January 6th Committee & january, committee, jan, impeachment, trial, j6, guilty, hearing, impeach, jan. \\
\addlinespace
Left-Wing Criticism & left, liberal, conservative, leftist, right, wing, lefty, radical, far, libs \\
\addlinespace
LGBTQ and Religious Issues & gay, trans, christian, religious, church, transgender, religion, marriage, catholic, pride \\
\addlinespace
Media Appearances and Events & watch, video, fox, news, share, listen, live, watching, interview, join \\
\addlinespace
Media Distrust and Misinformation & media, lie, fake, news, story, truth, propaganda, msm, narrative, lying \\
\addlinespace
Mental Health Stigma & bad, evil, worse, mental, brain, sad, absolutely, worst, sick, hypocrisy \\
\addlinespace
News and Press Releases & new, news, article, report, official, day, release, press, board, list \\
\addlinespace
Political Party Conflict & democrat, republican, party, dems, gop, democratic, dem, dnc, majority, moderate \\
\addlinespace
Presidential Approval Ratings & approval, rating, reagan, dementia, carter, johnson, jimmy, lincoln, uncle, pete \\
\addlinespace
Primary Elections and Campaigns & win, candidate, run, primary, running, campaign, winning, lose, lost, oz \\
\addlinespace
Racial Politics Discussion & white, black, racist, race, racism, blm, color, minority, racial, supremacist \\
\addlinespace
RINO and Establishment Critics & liz, rino, cheney, traitor, swamp, establishment, penny, bush, romney, mitt \\
\addlinespace
Russia-Ukraine War & russia, ukraine, war, putin, russian, military, nuclear, ukrainian, nato, sanction \\
\addlinespace
States and Governors & state, red, desantis, city, texas, florida, governor, blue, california, county \\
\addlinespace
Supreme Court Decisions & court, supreme, justice, judge, roe, scotus, wade, lawsuit, decision, sue \\
\addlinespace
Taxes and Government Spending & money, tax, billion, pay, dollar, million, paid, fund, irs, soros \\
\addlinespace
Trump Support and Presidency & trump, president, donald, maga, desantis, supporter, rally, potus, presidency, office \\
\addlinespace
US International Relations & china, chinese, taiwan, europe, saudi, european, eu, uk, north, minister \\
\addlinespace
Vulgar Political Insults & s***, guy, f***, f***ing, lol, oh, got, blame, man, yeah, stupid, idiot, aoc, clown, dumb, moron, troll, funny, joke, po \\
\bottomrule
\end{tabular}
\caption{Full set of topics with representative keywords.}
\label{app:topics_full}
\end{table}

\clearpage

\begin{figure*}[t]

\centering
\caption*{\textbf{Appendix E: Plots of other events statistically tested for temporal dynamics}}

\vspace{0.5em}

\begin{subfigure}{0.48\textwidth}
    \centering
    \includegraphics[width=\textwidth]{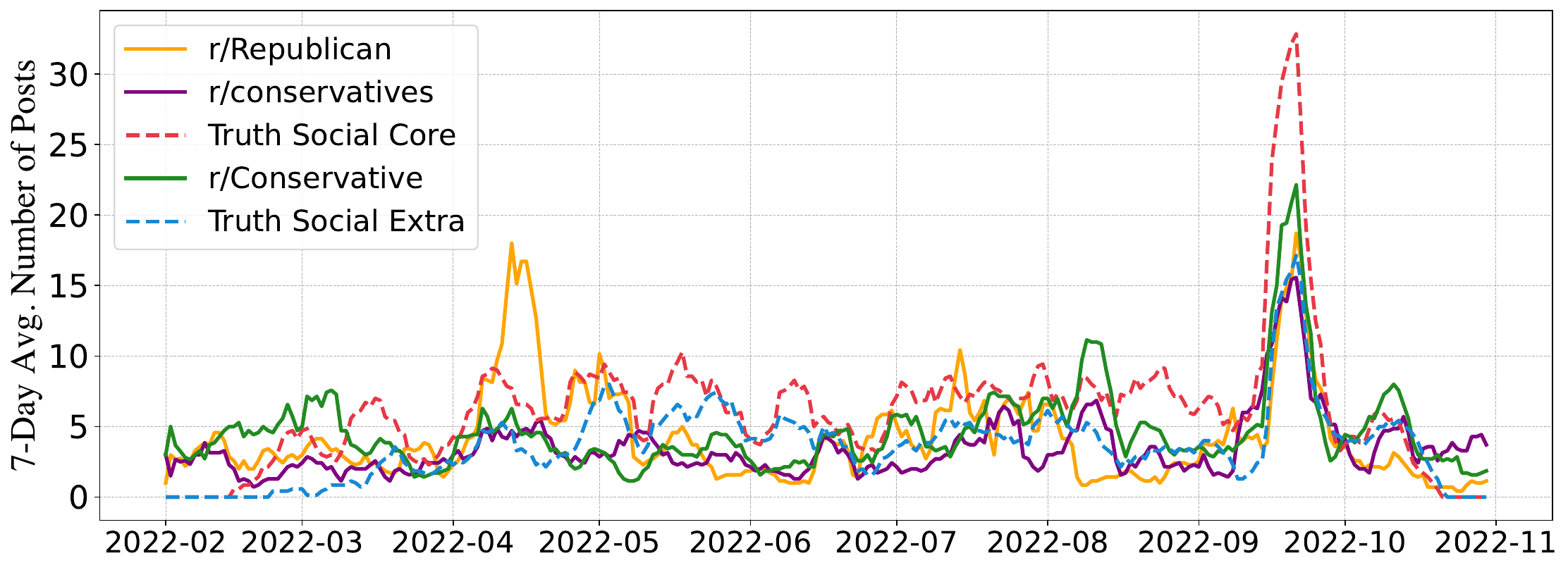}
    \caption{Border \& Immigration}
\end{subfigure}\hfill
\begin{subfigure}{0.48\textwidth}
    \centering
    \includegraphics[width=\textwidth]{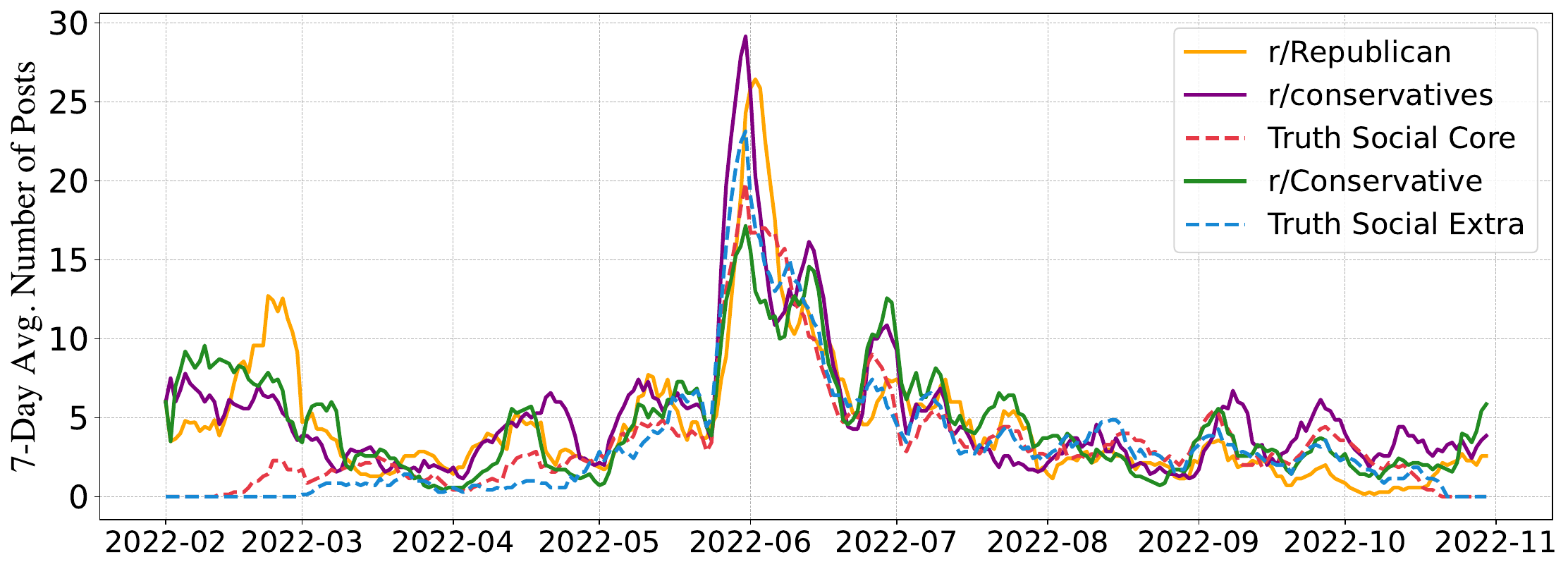}
    \caption{Gun Policy}
\end{subfigure}

\vspace{0.5em}

\begin{subfigure}{0.48\textwidth}
    \centering
    \includegraphics[width=\textwidth]{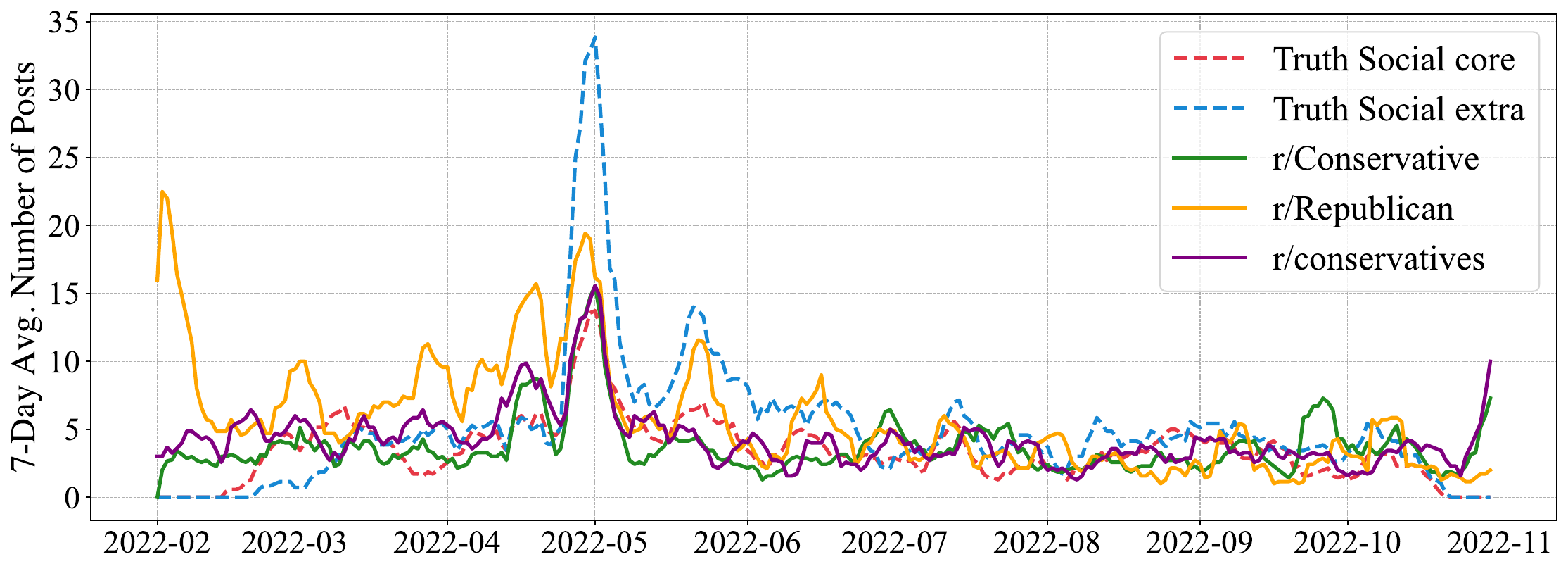}
    \caption{Big Tech Censorship}
\end{subfigure}\hfill
\begin{subfigure}{0.48\textwidth}
    \centering
    \includegraphics[width=\textwidth]{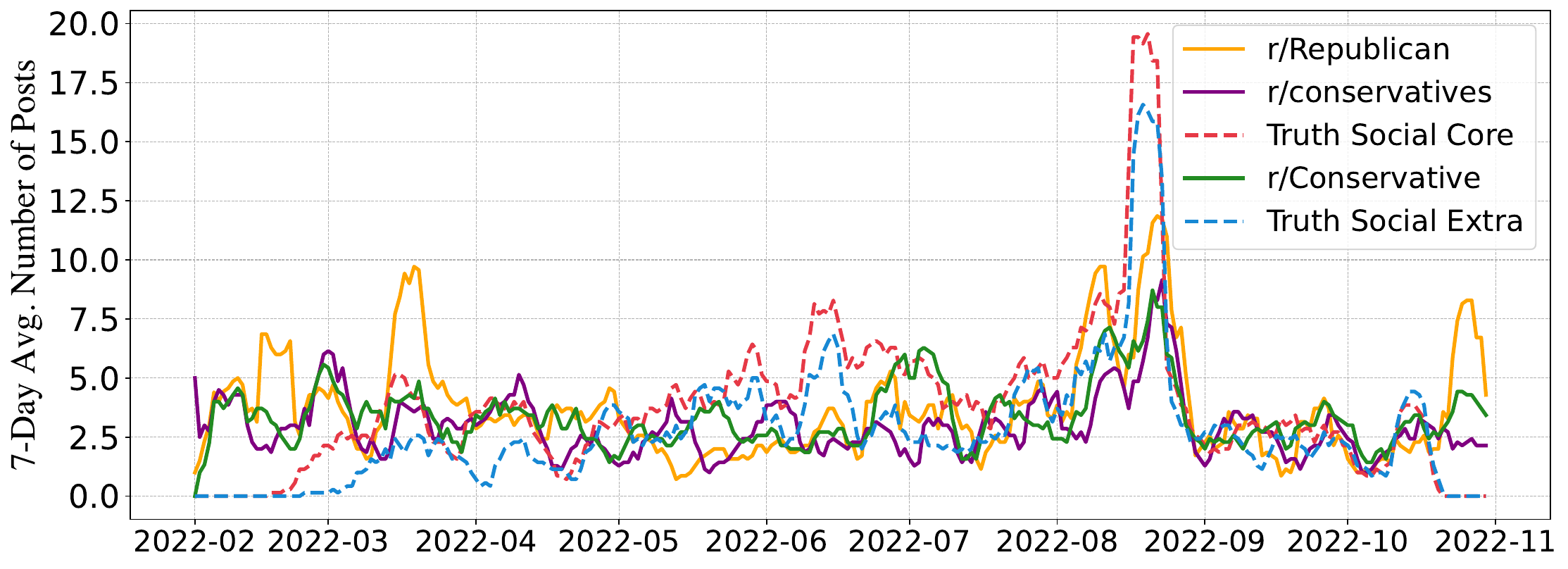}
    \caption{RINO and Establishment Critics}
\end{subfigure}

\label{fig:appendixE_temporal}
\end{figure*}

\begin{table*}[t]
\centering

\caption*{\textbf{Appendix F: Results of Time Series and Time Discontinuity Regression Models}}

\vspace{0.75em}

\begin{tabular}{ccccc}
\toprule
 & \multicolumn{2}{c}{User posting rate} & \multicolumn{2}{c}{Proportion of users} \\
\midrule
Predictors & Rate Ratios & 95\% Credible Intervals & Rate Ratios & 95\% Credible Intervals \\
\midrule
Intercept & 1.06 & 0.92--1.24 & 0.01 & 0.00--0.01 \\
source: r\_conservative & 0.89 & 0.78--1.01 & 0.24 & \textbf{0.16--0.38} \\
source: r\_conservatives & 1.06 & 0.92--1.22 & 0.26 & \textbf{0.17--0.40} \\
source: r\_republican & 1.15 & 1.01--1.31 & 0.96 & 0.78--1.17 \\
time\_diff & 0.97 & 0.92--1.02 & 1.20 & \textbf{1.01--1.42} \\
time: time1 & 1.17 & 0.99--1.37 & 1.72 & \textbf{1.09--2.77} \\
\midrule
Observations & 3250 &  & 160 \\
\bottomrule
\end{tabular}

\caption{Bayesian time discontinuity regression results for the topic of the Russian--Ukraine War.}
\label{tab:russia_posts}

\end{table*}

\begin{table*}[h]
    \centering
    \begin{tabular}{ccccc}
    \toprule
    & \multicolumn{2}{c}{User posting rate} & \multicolumn{2}{c}{Proportion of users}\\
    \midrule
    Predictors & Rate Ratios & 95\% Credible Intervals & Rate Ratios & 95\% Credible Intervals  \\
     \midrule
         Intercept	& 1.02	& 0.84 – 1.24 & 0.00	&0.00 – 0.01 \\

source: r\_conservative	& 0.90	&0.74 – 1.08 & 0.12&	\textbf{0.08 – 0.18} \\
source: r\_conservatives	& 0.95 &	0.77 – 1.16 &0.20	&\textbf{0.14 – 0.31} \\
source: r\_republican	&0.96	&0.78 – 1.17& 0.18&	\textbf{0.12 – 0.28 }\\
time\_diff &	0.96	&0.88 – 1.05 &1.02&	0.84 – 1.22  \\
time: time1	& 1.21	&0.93 – 1.55&1.21	&0.73 – 1.98 \\
\midrule
Observations	&1324 & &154 \\
\bottomrule
    \end{tabular}
    \caption{Bayesian time discontinuity regression results for the topic of Immigration.}
    \label{tab:immigration_posts}
\end{table*}

\begin{table*}[h]
    \centering
    \begin{tabular}{ccccc}
    \toprule
     & \multicolumn{2}{c}{User posting rate} & \multicolumn{2}{c}{Proportion of users}\\
    Predictors & Rate Ratios & 95\% Credible Intervals & Rate Ratios & 95\% Credible Intervals \\
     \midrule
     Intercept	& 0.99	& 0.82 – 1.21 &0.99	& 0.82 – 1.21 \\
source: r\_conservative	& 0.94	& 0.79 – 1.12  & 0.94	& 0.79 – 1.12\\
source: r\_conservatives&	1.06	&0.88 – 1.29  & 1.06&0.88 – 1.29 \\
source: r\_republican	& 1.15	& 0.96 – 1.38 & 1.15	& 0.96 – 1.38\\\
time\_diff &	0.98&	0.91 – 1.05 & 0.98&	0.91 – 1.05 \\
time: time1	& 1.13	& 0.90 – 1.42 	& 1.13	& 0.90 – 1.42 \\
\midrule
Observations	& 1625 &  & 165\\
\bottomrule
    \end{tabular}
    \caption{Bayesian time discontinuity regression results for the topic of Abortion.}
    \label{tab:abortion_posts}
\end{table*}

\begin{table*}[h]
    \centering
    \begin{tabular}{ccccc}
    \toprule
    & \multicolumn{2}{c}{User posting rate} & \multicolumn{2}{c}{Proportion of users}\\
    \midrule
    Predictors & Rate Ratios & 95\% Credible Intervals & Rate Ratios & 95\% Credible Intervals \\
     \midrule
     Intercept	& 1.36 &	1.24 – 1.50 & 0.01	 & 0.00 – 0.01 \\
source: r\_conservative	& 0.82	& 0.69 – 0.95 & 0.07 &	\textbf{0.05 – 0.10} \\
source: r\_conservatives &	1.03	& 0.91 – 1.16 &0.16 &	\textbf{0.11 – 0.24 } \\
source: r\_republican&	1.03	&0.89 – 1.20 &	0.13	&\textbf{0.08 – 0.19} \\
time\_diff	&1.03	&0.97 – 1.10 &	1.00	&0.85 – 1.19 \\
time: time1	&0.86&	0.72 – 1.02 &2.73	&\textbf{1.64 – 4.54} \\
\midrule
Observations	&  2907 &  & 159\\
\bottomrule
    \end{tabular}
    \caption{Bayesian time discontinuity regression results for the topic of FBI and DOJ Investigations.}
    \label{tab:fbi_posts}
\end{table*}

\begin{table*}[h]
    \centering
    \begin{tabular}{ccccc}
    \toprule
     & \multicolumn{2}{c}{User posting rate} & \multicolumn{2}{c}{Proportion of users}\\
    \midrule
    Predictors & Rate Ratios & 95\% Credible Intervals & Rate Ratios & 95\% Credible Intervals \\
     \midrule
     Intercept	& 1.36 &	1.24 – 1.50 & 0.01	&0.00 – 0.01 \\
source: r\_conservative	& 0.82	& 0.69 – 0.95 &0.09	&\textbf{0.06 – 0.14 } \\
source: r\_conservatives &	1.03	& 0.91 – 1.16 &0.17	&\textbf{0.11 – 0.26 }\\
source: r\_republican&	1.03	&0.89 – 1.20  &0.19	&\textbf{0.13 – 0.29}\\\
time\_diff	&1.03	&0.97 – 1.10 &0.99	&0.83 – 1.18 \\
time: time1	&0.86&	0.72 – 1.02  &	0.79	&0.48 – 1.33 \\
\midrule
Observations	& 1116  & & 163\\
\bottomrule
    \end{tabular}
    \caption{Bayesian time discontinuity regression results for the topic of RINO and Establishment Critics.}
    \label{tab:rino_posts}
\end{table*}

\begin{table*}[h]
    \centering
    \begin{tabular}{ccccc}
    \toprule
     & \multicolumn{2}{c}{User posting rate} & \multicolumn{2}{c}{Proportion of users}\\
     \midrule
    Predictors & Rate Ratios & 95\% Credible Intervals & Rate Ratios & 95\% Credible Intervals \\
     \midrule
 Intercept	& 1.07	& 0.99 – 1.17 & 0.01 &	0.00 – 0.01\\
source: r\_conservative	& 0.94	& 0.85 – 1.05 &	0.09&	\textbf{0.07 – 0.10} \\
source: r\_conservatives	& 0.99	&0.90 – 1.10 &0.17&	\textbf{0.14 – 0.21} \\
source: r\_republican	&1.00	&0.91 – 1.10  &0.19&	\textbf{0.16 – 0.22} \\
time\_diff	&1.00	& 0.996 – 1.003 	&0.98	&\textbf{0.97 – 0.99} \\
\midrule
Observations &5721 & & 1005 & \\
\bottomrule
    \end{tabular}
    \caption{Bayesian time series regression results for the topic of Big Tech Censorship and Free Speech.}
    \label{tab:tech_posts}
\end{table*}

\begin{table*}[h]
    \centering
    \begin{tabular}{ccccc}
    \toprule
     & \multicolumn{2}{c}{User posting rate} & \multicolumn{2}{c}{Proportion of users}\\
     \midrule
    Predictors & Rate Ratios & 95\% Credible Intervals & Rate Ratios & 95\% Credible Intervals \\
     \midrule
 Intercept	& 1.08	& 0.98 – 1.20 & 0.00 &	0.00 – 0.01\\
source: r\_conservative	& 0.95	& 0.86 – 1.05 &	0.17&	\textbf{0.10 – 0.27} \\
source: r\_conservatives	& 1.03	&0.93 – 1.14 &0.24&	\textbf{0.15 – 0.40} \\
source: r\_republican	&1.06	&0.95 – 1.18  &0.21&	\textbf{0.13 – 0.34} \\
time\_diff	&1.00	& 0.995 – 1.003 	&0.90	&0.75 – 1.07 \\
\midrule
Observations &5375 & & 114 & \\
\bottomrule
    \end{tabular}
    \caption{Bayesian time series regression results for the topic of Gun Policy.}
    \label{tab:gun_policy}
\end{table*}

\begin{table*}[h]
    \centering
    \begin{tabular}{ccccc}
    \toprule
     & \multicolumn{2}{c}{User posting rate} & \multicolumn{2}{c}{Proportion of users}\\
     \midrule
    Predictors & Rate Ratios & 95\% Credible Intervals & Rate Ratios & 95\% Credible Intervals \\
     \midrule
 Intercept	& 1.05	& 0.90 – 1.25 & 0.02 &	0.01 – 0.03\\
source: r\_conservative	& 0.96	& 0.82 – 1.14 &	0.24&	\textbf{0.16 – 0.37} \\
source: r\_conservatives	& 1.05	&0.86 – 1.28 &0.28&	\textbf{0.18 – 0.43} \\
source: r\_republican	&1.18	&\textbf{1.01 – 1.39}  &0.43&	\textbf{0.29 – 0.64} \\
time\_diff	&0.99	& 0.91 – 1.06 	&2.70	&\textbf{2.08 – 3.49} \\
time: time1	&1.02	& 0.85 – 1.23 	&0.59	&0.36 – 1.01 \\
time\_diff:time1	&  	&  	&0.18	&\textbf{0.13 – 0.26} \\
\midrule
Observations &1950 & & 164 & \\
\bottomrule
    \end{tabular}
    \caption{Bayesian time discontinuity regression results for the topic of Energy Crisis Policy.}
    \label{tab:energy_policy}
\end{table*}

\end{document}